\documentclass[sigconf]{acmart}

\usepackage[utf8]{inputenc}
\usepackage{graphicx}
\usepackage{xcolor}
\usepackage{color}
\usepackage{xspace}
\usepackage{hyperref}
\usepackage{booktabs}
\usepackage{tabularx}
\usepackage{tikz}
\usepackage{color}
\usepackage{soul}
\usepackage{amsmath}
\usepackage{mathtools}
\usepackage[linesnumbered,ruled, vlined]{algorithm2e}
\usepackage{cleveref}
\usepackage{xifthen}
\usepackage{pifont}
\usepackage{caption}
\usepackage{subcaption}
\usepackage[inline]{enumitem}
\usepackage{multirow}
\usepackage{makecell}
\usepackage{chemformula}
\usepackage{todonotes}
\usepackage{siunitx}
\usepackage{pgf}

\usepackage[group-separator={,},group-minimum-digits=4]{siunitx}
\usepackage[super]{nth}
\usepackage{chemformula}
\usepackage{eurosym}

\usepackage{amsmath,amsfonts}
\usepackage{algorithmic}
\usepackage{graphicx}
\usepackage{adjustbox}
\usepackage[inline]{enumitem}
\usepackage{textcomp}
\usepackage{lipsum}
\usepackage{xcolor}
\usepackage{tikz}
\usepackage{float}
\usepackage{placeins}
\usepackage{listings}
\usepackage{pifont}
\PassOptionsToPackage{hyphens}{url}\usepackage{hyperref}
\usepackage{cleveref}
\usepackage{tabularray}
\usepackage{pifont}
\usepackage{tabularx}
\usepackage{stfloats}

\definecolor{darkred}{rgb}{0.5,0,0}
\definecolor{darkgreen}{rgb}{0,0.5,0}
\definecolor{darkblue}{rgb}{0,0,0.5}

\newcommand{\circled}[1]{%
    \tikz[baseline=(char.base)]{
        \node[shape=circle, fill=black, text=white, inner sep=0.5pt, font=\small] (char) {#1};
    }%
}

\def\BibTeX{{\rm B\kern-.05em{\sc i\kern-.025em b}\kern-.08em
    T\kern-.1667em\lower.7ex\hbox{E}\kern-.125emX}}


\copyrightyear{2026}
\acmYear{2026}
\setcopyright{cc}
\setcctype{by}

\makeatletter
\let\old@copyrightpermission\@copyrightpermission
\def\@copyrightpermission{%
  A subset of this article has been accepted for publication, following peer review, at ACM Computing Frontiers 2026. The open-science artifacts have received all ACM reproducibility badges from the reviewers and are publicly available.\par\vspace{4pt}%
  \old@copyrightpermission
}
\makeatother

\acmConference[Technical Report]{Technical Report}{Homonym of article at Computing Frontiers}{2026}
\acmBooktitle{}
\acmDOI{}
\acmISBN{}

\title{M3SA: Exploring Datacenter Performance and Climate-Impact with Multi- and Meta-Model Simulation and Analysis \\ \textcolor{darkgray}{-- Extended Technical Report --}}

\author{Radu Nicolae}
\email{R.Nicolae@vu.nl}
\orcid{0009-0007-0318-9266}
\affiliation{%
  \institution{Vrije Universiteit Amsterdam}
  \city{Amsterdam}
  \country{The Netherlands}
}

\author{Dante Niewenhuis}
\email{D.Niewenhuis@vu.nl}
\orcid{0000-0002-9114-1364}
\affiliation{
  \institution{Vrije Universiteit Amsterdam}
  \city{Amsterdam}
  \country{The Netherlands}
}

\author{Sacheendra Talluri}
\email{S.Talluri@vu.nl}
\orcid{0000-0003-3461-4919}
\affiliation{%
  \institution{Vrije Universiteit Amsterdam}
  \city{Amsterdam}
  \country{The Netherlands}
}

\author{Alexandru Iosup}
\email{A.Iosup@vu.nl}
\orcid{0000-0001-8030-9398}
\affiliation{%
  \institution{Vrije Universiteit Amsterdam}
  \city{Amsterdam}
  \country{The Netherlands}
}

\begin{abstract}

Datacenters are vital to our digital society, but consume a considerable fraction of global electricity and demand is projected to increase. 
To improve their sustainability and performance, we envision that simulators will become primary decision-making tools. 
However, 
and unlike other fields focusing on key societal infrastructure such as waterworks and mass transit, 
datacenter simulators do not yet combine multiple independent models into their operation and thus suffer from issues associated with singular models, such as specialization, and lack of adaptability to operational phenomena. 
To address this challenge, we propose M3SA, a datacenter simulation and analysis framework that uses discrete-event simulation to predict, for each model, the impact on climate and performance under various realistic datacenter conditions, and then combines these predictions. 
We design 
an architecture for simulating multiple concurrent models (\textit{Multi-Model}), 
a technique for integrating the results of multiple models into a \textit{Meta-Model}, and 
a procedure for quantifying Meta-Model accuracy.
Through experiments with an M3SA prototype, we show that 
(i) M3SA can reproduce and enhance peer-reviewed experiments;
(ii) M3SA can predict operational phenomena (e.g., failures) of datacenters, running fundamentally different workload traces;
(iii) M3SA enables various types of what-if and how-to analysis, such as how to configure CO2-aware migration over yearly energy-production patterns.
M3SA has been integrated into the open-source simulator OpenDC and is available at: ~\url{https://github.com/atlarge-research/opendc-m3sa}.

\end{abstract}


\begin{document}


\maketitle

\section{Introduction} \label{sec:introduction}

Our society and economy are increasingly dependent on digital services. Correspondingly, new datacenters are being built, and existing datacenters are scaled up~\cite{DBLP:conf/sc/AndreadisVMI18, market:IDC23, market:IDC24AI, market:Gartner24}. Before building or expanding, operators use simulators to project infrastructure performance and climate impact~\cite{graphmassivizer, DBLP:conf/mascots/RenWUS12, DBLP:journals/tpds/TuliPSCJ22}. For example, in the EU research project Graph Massiziver, 
simulators predict speedup, failure cost, energy consumption, and CO2 emissions for massive-scale infrastructure~\cite{conf/cloudsummit/ProdanKB+22}. 
Although datacenter simulation remains challenging in general, for example, due to the unpredictable behavior of 
datacenter software and hardware ecosystems~\cite{DBLP:conf/icdcs/IosupUVAEHTBT18, DBLP:conf/hotos/BronsonACZ21} and the continuous emergence of new devices and applications~\cite{market:Gartner24,market:IDC24AI}, in this work, we focus on the model at the core of simulation. 
Current simulators, such as the open-source OpenDC~\cite{DBLP:conf/ccgrid/MastenbroekAJLB21}, CloudSim~\cite{DBLP:journals/spe/CalheirosRBRB11,DBLP:journals/spe/HewageIRB24}, and SimGrid~\cite{DBLP:conf/ccgrid/Casanova01,DBLP:journals/fgcs/McDonaldDWSC24}, already support \textit{single-model simulation}--the simulation is based on a single integral model, where each component or phenomenon is modelled with a partial model, e.g., one for CPUs and one for the storage subsystem. 
However, as this work showcases by quantifying single-model error in \Cref{sec:experiments:exp1}, \textit{single} models are often trained for specific scenarios and are prone to errors when encountering edge cases and new scenarios.
Addressing this gap, in this work, we design, prototype, and evaluate the M3SA (read as [mesa]) framework for
\textit{multi-model} (and \textit{meta-model}) simulation and analysis of datacenters.

\begin{figure}[t]
    \centering
    \includegraphics[width=\linewidth]{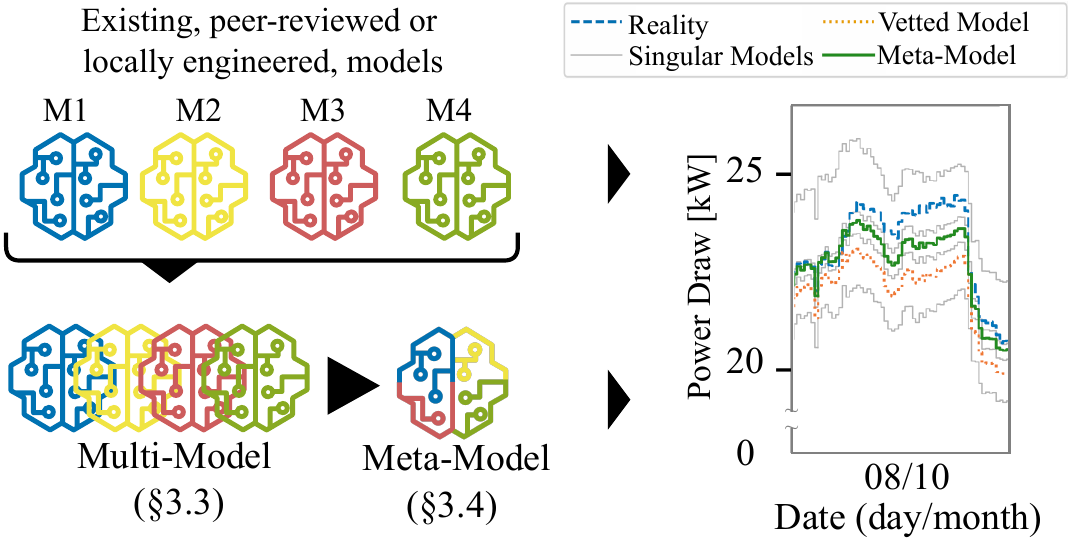}
    \caption{M3SA Multi-Model simulation uses multiple models simultaneously. M3SA Meta-Model integrates all results. \protect\Cref{fig:exp1:A-B-C:results} depicts a specific scenario.}
    \label{fig:introduction:float-1}
    \vspace*{-0.65cm}
\end{figure}

Albeit essential, we argue that single integral (\textit{singular}) models are insufficient for reliable predictions -- a singular model makes accurate predictions for the context it has been trained for, and often fails to capture real-life edge-case scenarios~\cite{covid19ForecastHubUS, environmentMultiModel, DBLP:conf/ccgrid/MastenbroekAJLB21}. Moreover, singular models are calibrated to ideal scenarios and may lack adaptability to different \textit{operational phenomena}, such as spikes in demand or equipment failures. In this work, we posit that combining multiple models and contrasting their predictions for the same simulated scenario can retain the strengths of each model and alleviate biases. The community leverages diverse modelling approaches trained or validated on diverse datasets~\cite{DBLP:conf/im/FilhoOMIF17, DBLP:conf/iccS/SilvaOCTDS19, DBLP:conf/ccgrid/MastenbroekAJLB21,  DBLP:journals/tsusc/HuangCPZA24,DBLP:journals/fgcs/ArdebiliABB24,DBLP:journals/tpds/AksarSSALBKEC24}---each an opportunity to include multi-model simulation.

\Cref{fig:introduction:float-1} illustrates our approach, as a high-level abstraction~(left) and through a real-world experiment (right, sample from \S\ref{sec:experiments:exp1}).
Here, M3SA consists of four singular and peer-reviewed models. 
Firstly, M3SA assembles a \textit{(i) Multi-Model}, by leveraging predictions of individual into a unified entity (e.g., plot, dataframe).
Then, M3SA leverages a \textit{(ii) Meta-Model}, which predicts by aggregating predictions of previously assembled models.
\Cref{fig:introduction:float-1}~(right) showcases the Meta-Model alleviating individual biases and better aligning with the measured reality. In this experiment, the Meta-Model delivers better accuracy than a state-of-the-art, peer-reviewed model~\cite{DBLP:conf/wosp/NiewenhuisTIM24}.

Simulating datacenters using multiple models is a novelty in computer systems. 
However, other computer-science-related fields already adopt variations of multi- and meta-models (e.g., ensemble learning and multi-model combination techniques in machine learning (ML) and statistical modelling). Conceptually closest to the M3SA \textit{Meta-Model}, we identify bootstrap aggregating (bagging), a state-of-the-art in ML for improving accuracy through prediction aggregation~\cite{DBLP:journals/ml/Breiman96b}. Beyond computer science, large-scale, coarse-grained, simulations in weather and climate simulations~\cite{schunk2016space-multimodel-weather, environmentMultiModel}, and small-scale, detailed simulations in ecology~\cite{ecologyMultiModel} already use instruments that rely on multiple models, calibrated and adjusted for the specific needs of their scientific field.

In contrast, to obtain meaningful predictions through multi-model simulation of datacenter ecosystems, medium-scale models must combine higher-level abstraction with detailed operational models for specific devices and applications, and address various operational phenomena. This is fundamentally different from other-scale sciences~\cite{allen2019hierarchy}. Currently, none of the open-source datacenter simulators supports multi-model simulation, including the highly cited and much-used
CloudSim~\cite{DBLP:journals/spe/CalheirosRBRB11,DBLP:journals/spe/HewageIRB24}, SimGrid~\cite{DBLP:conf/ccgrid/Casanova01,DBLP:journals/fgcs/McDonaldDWSC24},
OpenDC~\cite{DBLP:conf/ccgrid/MastenbroekAJLB21}, and
GDCSim~\cite{DBLP:conf/green/GuptaGBAMV11}.
One of the possible drawbacks of multi-model simulation, which we test for explicitly in this work, is the computational overhead in simulation, over traditional -model approaches; we quantify this overhead in \Cref{sec:experiments:exp1}.

In this work, we address the open challenge
of \textit{leveraging multiple models and integrating their results into a Meta-Model}
with a three-fold contribution:
\begin{enumerate}[label=\textbf{C\arabic*}]    
    \item \label{introduction:c1} (\textbf{Design}) We design M3SA, the first \underline{m}ulti- and \underline{m}eta-\underline{m}odel \underline{s}imulator for d\underline{a}tacenters~(\Cref{sec:m3sa}), and design the Multi- and Meta-Model components to work with generic discrete-event simulators.
    Overall, M3SA allows the same component to be modelled in different ways and systematically combine predictions of distinct, independently run, singular models.

    \item \label{introduction:c2} (\textbf{Experimentation}) We prototype M3SA and embed into OpenDC~\cite{DBLP:conf/ccgrid/MastenbroekAJLB21}, a peer-reviewed, state-of-the-art datacenter simulator. We further equip M3SA with 8 commonly used, peer-reviewed models, and conduct three large-scale, trace-driven datacenter experiments to:
    (i) Reproduce a peer-reviewed, open-science datacenter experiment~\cite{DBLP:conf/wosp/NiewenhuisTIM24} (\S\ref{sec:experiments:exp1}); 
    (ii) Analyze and compare the impact of workload kind, i.e., scientific~\cite{DBLP:journal/nature/BorghesiSDMBGMCGCBBB23} vs. business-critical~\cite{DBLP:conf/ccgrid/ShenBI15}, on datacenter performance and CO2-emissions~(\S\ref{sec:experiments:exp2}); and
    (iii) Analyze the impact of EU-region on the performance and CO2-emissions of migrated workload~(\S\ref{sec:experiments:exp3}).

    \item \label{introduction:c3} (\textbf{Open science}) 
    We contribute to open science all the software and data~\cite{site:trace-archive} created in this project.
    For the last experiment, we collect public CO2-emissions data across all EU countries from ENTSO-E, align the timestamps, compute carbon-intensity data from it, and release the resulting FAIR dataset.
    Last, we make available the complete reproducibility capsule: \url{https://anonymous.4open.science/r/m3sa-hcp/}.

\end{enumerate}

\section{Background: Simulation and System Model} \label{sec:background}
\begin{figure}[t]
    \centering
    \includegraphics[width=0.99\linewidth]{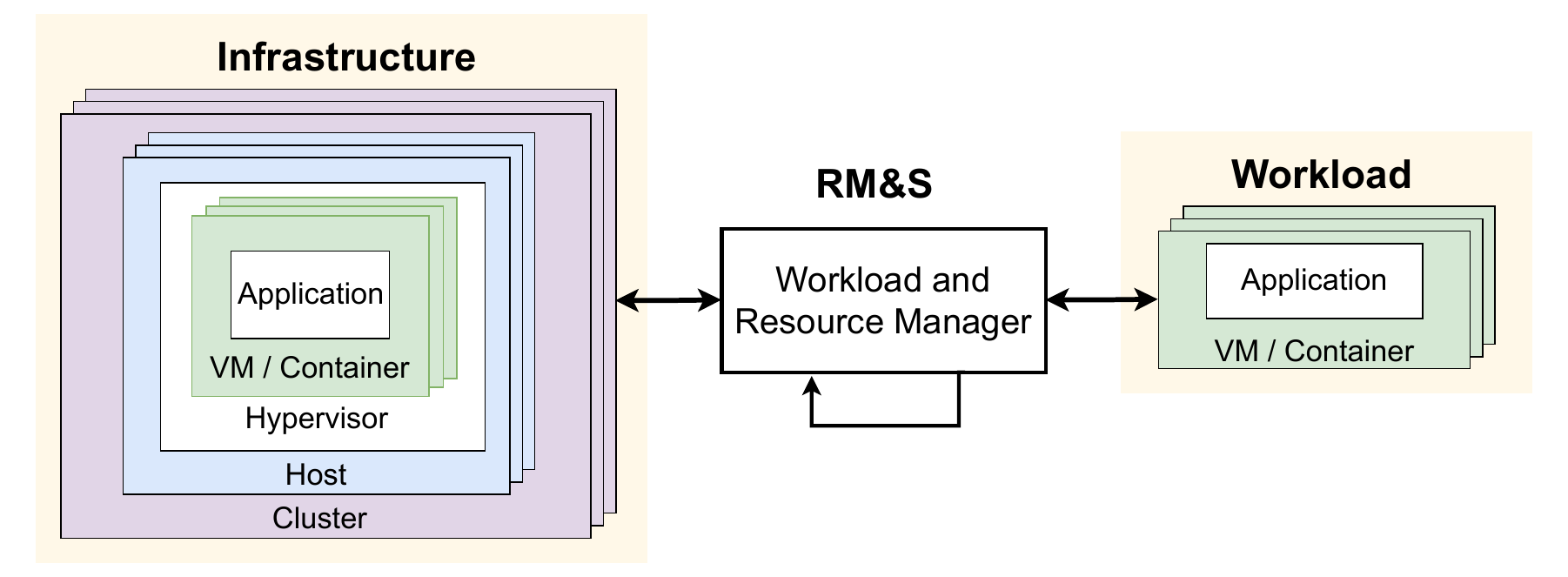}
    \vspace*{-0.20cm}
    \caption{System model for datacenter operation adapted from~\cite{DBLP:journals/tpds/AndreadisMBI22}. Simulation enables fine-grained datacenter operational monitoring used by stakeholders in decision-making processes.}
    \vspace*{-0.65cm}
    \label{fig:f2-background-image}
\end{figure}

We present background on datacenter simulators and a basic system model, illustrated by \Cref{fig:f2-background-image}. Combining the two, we present elements of prediction models and metrics widely used in the community and useful for understanding this work.

\begin{figure*}[t]   
    \includegraphics[width=\linewidth]{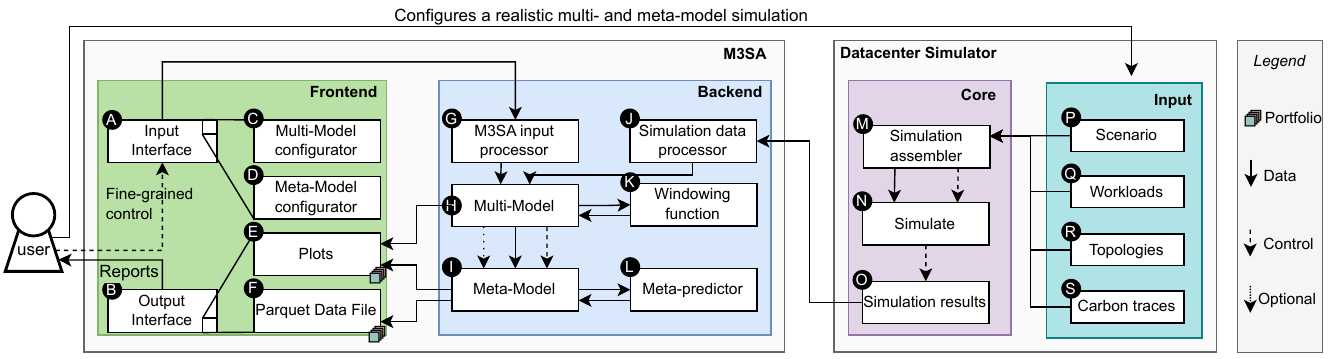}
    \vspace*{-0.5cm}
    \caption{Architecture of M3SA, a Multi- and Meta- Model Simulation Analyzer, integrated with a black-boxed simulator.}
    \label{fig:f3-m3sa-overview}
    \vspace*{-0.15cm}
\end{figure*}

\noindent\textit{``Simulation is defined as the imitation of the operation of a system or real-world process over time, and in many cases, manufacturing provides one of the most important applications of simulation``}~\cite{DBLP:conf/wsc/LeeMS03}. Simulation provides datacenter stakeholders with operational insights into the behavior of infrastructure running under a given workload. 

\Cref{fig:f2-background-image} illustrates our basic system model, which is an abstraction of many real-world operations. \textit{Workloads} contain tasks operated on \textit{infrastructure}, which includes physical machines, virtual machines~(VM), and containers; a complex \textit{resource management and scheduling (RM\&S) system} is continuously matching workloads to infrastructure, subject to optimization goals such as energy use (power draw) and CO2-emission reduction. \textit{Traces} represent recordings of real-world events, capturing detailed operational data of the infrastructure under given workload(s); traces provide a granular view of resource usage, essential for driving simulations or replaying real-world scenarios.

\textit{Predictive models} in large-scale computer systems are empirical prediction systems that analyze, combine, and compute various input elements (often, traces) to produce fine-grained output predictions~\cite[\S 1]{modsim:book/ZaraiN15:orig}. We use such models to predict how real-world workloads run on ICT infrastructure, under various operational phenomena, as modeled by users. In this work, we use three types of models: \begin{enumerate*}[label=(\textit{\roman*})]
    \item \textit{Energy models} predict the amount of energy the datacenter consumes from the power grid;
    \item \textit{CO2 emission models} predict the carbon intensity (CO2 emitted per unit of energy, measured in $gCO_{2}$) of the datacenter under the simulated workload;
    \item \textit{Failure models} simulate task failures according to a failure trace. Without such traces, failure models generate task failures with arbitrary but realistic occurrence timestamps and recovery periods. 
\end{enumerate*}

Our technical report attached to this work (appendices \ref{sec:appendix:a}-\ref{sec:appendix:d}) details each simulation model used in this work.

We use \textit{metrics} for system monitoring and accuracy measurement. For system monitoring, we use two main metrics: \textit{energy consumption} and \textit{carbon intensity}, measured in international system units. We define and measure the models' accuracy against real-world data (i.e., measured reality) using \textit{Mean Percentage Absolute Error (MAPE)}, expanded in \Cref{sec:mesa:meta-model:validation}.

\section{Design of M3SA: A tool for Multi- and Meta-Model Simulation and Analysis} \label{sec:m3sa} 

In this section, we synthesize requirements and design a novel approach for datacenter simulation. We propose M3SA, a system that enables simulation using \textit{multiple} models.

\subsection{Requirement Analysis} \label{sec:m3sa:requirements-analysis}
\begin{enumerate}[label=\textbf{(FR\arabic*)},leftmargin=0pt,itemindent=3em]
    \item \label{m3sa:fr1} \textbf{Leverage multiple models into a unified tool}: The system must enable combining multiple (existing) models into a unified tool. It should enable visualizing their predictions, independently or combined, thus increasing the explainability of predictions, and pointing out idiosyncratic errors or biases.
    \item \label{m3sa:fr2} \textbf{Predict by integrating multiple models}: A \textit{Meta-Model}, able to aggregate singular-models' predictions, at a user-selected granularity, to alleviate errors and biases of idiosyncratic models, thus strengthening the credibility of the predictions.
    \item \label{m3sa:fr3} \textbf{Open-source tool:} The system should be open-sourced. The system should contribute to open science, both as coupled with a peer-reviewed, vetted, and state-of-the-art simulator and decoupled, as an independent scientific tool.
\end{enumerate}

\begin{enumerate}[label=\textbf{(NFR\arabic*)},leftmargin=0pt,itemindent=3.7em]
    \item \label{m3sa:nfr1} \textbf{Provide in-meeting or same-day simulation results}: 
    Multiple models can be run in parallel, reducing the simulation time to a matter of minutes or less\footnote{Complex datacenter experiments using OpenDC can be used as a reference point. They are reported to take seconds to minutes~\cite{DBLP:conf/ccgrid/MastenbroekAJLB21}.}.
    The analysis part of the system should only add a fraction to the simulation time itself, and not more than double it.
    \item \label{m3sa:nfr2} \textbf{Highly accurate predictions:} The Meta-Model should predict with higher accuracy than the average singular model employed in the simulation. Equivalently, the error rate of the Meta-Model should be lower than the error rate of the average model.
    \item \label{m3sa:nfr3} \textbf{Operation at real-world scale}: All FRs and NFRs should be met for \textit{at least} 8 singular models, on input traces of \textit{at least} 200,000 samples per model, equivalent to 2 years at the industry/science de facto sampling rate of 5 minutes~\cite{DBLP:conf/ccgrid/MastenbroekAJLB21, DBLP:journals/tpds/AndreadisMBI22}. Although a more general aim would be to support as many models as possible, as the experimental results in \S\ref{sec:experiments:exp1} indicate, already 8 such models, combined, can deliver good results when compared with state-of-the-art hand-tuned models.
\end{enumerate}

\subsection{Overview of the M3SA Architecture} \label{sec:m3sa:overview}

We design M3SA to be capable of operating coupled or decoupled from a datacenter simulator. \Cref{fig:f3-m3sa-overview} depicts an overview of the system's architecture, in which M3SA extends a black-boxed simulator \ref{m3sa:fr3}.
In our experience, the main effort to couple M3SA to a specific simulator is put in matching the simulator output with the M3SA input. 
We test this by first designing and prototyping M3SA, and then coupling it with the state-of-the-art simulator OpenDC~\cite{DBLP:conf/ccgrid/MastenbroekAJLB21, DBLP:conf/sc/AndreadisVMI18}. 
OpenDC is a peer-reviewed, open-source, discrete-event simulator with simple interfaces, and over 5 years of development and operation~\cite{DBLP:conf/ccgrid/MastenbroekAJLB21, DBLP:conf/sc/AndreadisVMI18}.

\begin{figure*}[ht]
    \centering
    \includegraphics[width=0.32\linewidth]{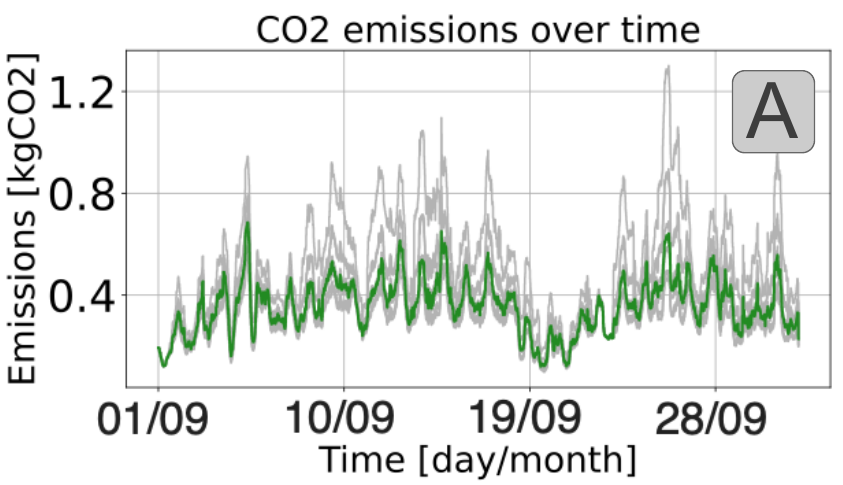}
    \hfill
    \includegraphics[width=0.32\linewidth]{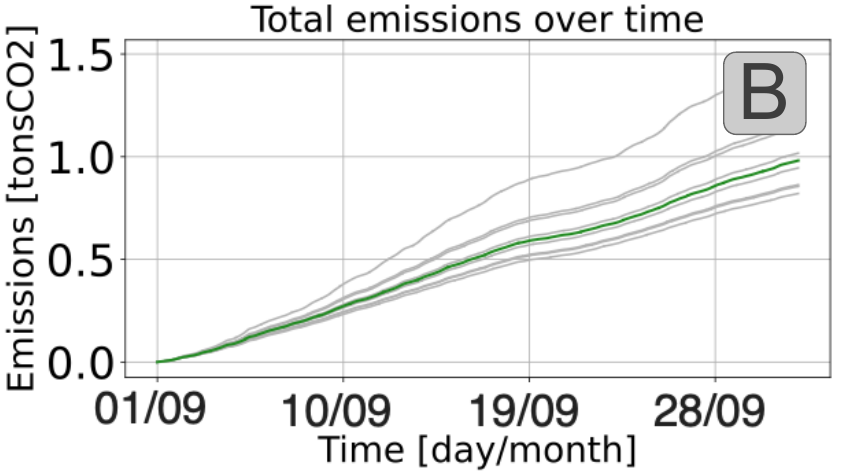}
    \hfill
    \includegraphics[width=0.32\linewidth]{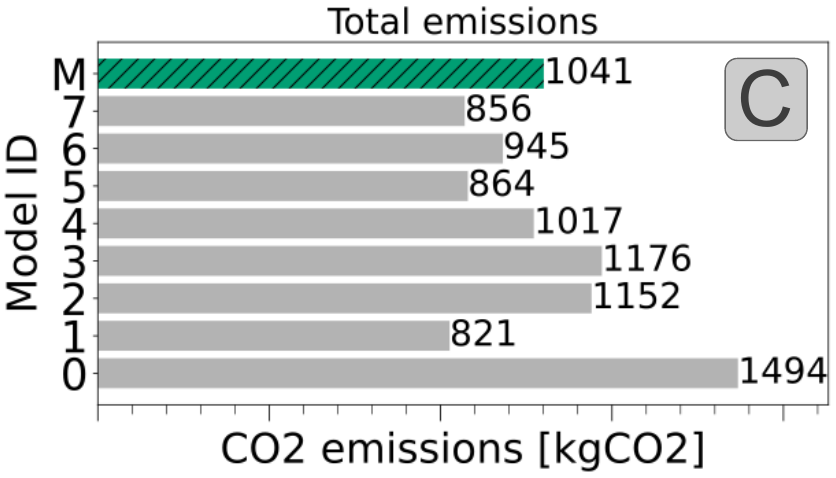}
    \vspace*{-0.35cm}
    \caption{Sample M3SA plots. We represent singular models in gray and Meta-Models in green. In (A),(B): vertical axis depicts simulated metric, horizontal depicts time. In (C): vertical axis depicts model identifier, horizontal depicts cumulative totals.}
    \label{fig:results:multimodelcomponent}
    \vspace*{-0.35cm}
\end{figure*} 

M3SA adopts the architecture we define as ``Simulate First, Compute Later'' (SFCL), where M3SA operates only after the simulator successfully predicts and the \textit{simulation results}~(\circled{O}) are available. This architecture couples M3SA to the simulator as an external layer. We also considered the alternative architecture of ``Compute While Simulating'' (CWS), where M3SA would be deeply embedded into the simulation infrastructure. Comparing SFCL and CWS architectures, we observe a tradeoff between the depth of M3SA embedding \ref{m3sa:fr3} and the performance of the overall system \ref{m3sa:nfr1} - the more embedded M3SA is, the better the overall performance, yet the more complex it becomes to port M3SA from one simulator to another. We choose the SFCL architecture, thus allowing simple adoption of M3SA; to meet the performance requirements, we focus on reducing redundancies and following good engineering practices.

In the SFCL architecture, the user interacts with the system through the \textit{Input Interface}~(component \circled{A}) and \textit{Output Interface}~(\circled{B}) interfaces \ref{m3sa:fr3}. The M3SA process begins from the input, where the users configure the \textit{Multi-Model}~(\circled{C}) and the \textit{Meta-Model}~(\circled{D}). The simulation process, triggered and controlled by the M3SA backend, happens between~\circled{M}-\circled{S}: the system sets up a simulation based on user-input, simulates, and centralizes predictions. The simulation block~\circled{M}-\circled{S} corresponds to how discrete-event simulators commonly used in the field actually operate, and closely matches the OpenDC~\cite{DBLP:conf/ccgrid/MastenbroekAJLB21} and CloudSim~\cite{DBLP:journals/spe/HewageIRB24} architectures; in particular, the simulation assembler~(\circled{M}) is where singular models are typically defined in current experiments.

Further, the M3SA backend retrieves the predictions made by the simulation infrastructure, one per singular model, and assembles the Multi-Model \ref{m3sa:fr1}. To avoid redundant computation, the \textit{Meta-Model} predicts using already leveraged and processed data from the Multi-Model \ref{m3sa:fr2}, \ref{m3sa:nfr1}. Lastly, the system plots the predictions of the Multi- and Meta-Model~(\circled{E}) and outputs the Meta-Model predictions to a lossless-compressed data file~(\circled{F}).

\subsection{Design of the Multi-Model Component} \label{sec:m3sa:multi-model:design}

In this section, we define explainability, and elaborate on the \textit{Multi-Model} and \textit{Plots} components (\Cref{fig:f3-m3sa-overview} \circled{H} and \circled{E}, respectively).
The M3SA multi-model approach helps compare the outputs singular models, thus improving users' understanding of the predictions and increasing explainability \ref{m3sa:fr1}.
It also enables meta-modeling, as detailed in \Cref{sec:m3sa:meta-model:design}.

We define \textit{explainability} as the user's understanding of the possible behavior, limitations, and biases of the system under test, as predicted by relevant, peer-reviewed or locally vetted, models.
Better explainability is achieved when multiple models are available, and when at least the ranges of acceptable predictions, and also the detailed singular predictions, are available and contrasted across the models, enabling specialists to better reason about the problem\footnote{In many scientific domains, e.g., in weather and climate modeling~\cite{schunk2016space-multimodel-weather, environmentMultiModel}, multiple instrumental observations (traces) and multiple models are used to provide evidence, toward building sufficient confidence in specific outcomes.}. Ideal explainability both matches measurable reality and links real-world values to predictions made by one or several models, thereby enabling specialists to explain wherever evidence exists about which model gives better predictions and how close those predictions are to instrumental observations (traces)---proxy to reality. 

The Multi-Model component~(\circled{H}) leverages multiple models into a unified tool.
Within this component, predictions of singular models, which M3SA has already produced through its Datacenter Simulator backend, are centralized; the user can select the form of data representation. To increase performance and alleviate redundancy, the system reads from the input data file only the metrics selected by the user (typically, at configuration time) and discards irrelevant data. The Multi-Model process then filters prediction data from singular models with a user-selected windowing function, which minimizes the dataset for better performance \ref{m3sa:nfr1} and also removes noise for better visual comprehension;
we expand on the concept of windowing in~\Cref{sec:m3sa:multi-model:window-size}. After completion, the Multi-Model is ready for further operations, such as Meta-Model computation or plotting \ref{m3sa:fr2}.

The Plots component~(\circled{E}) uses the Multi-Model results to produce, guided by typical user-given configuration, visual outcomes as in~\Cref{fig:results:multimodelcomponent}.
M3SA supports analyzing how metrics evolve over time, enabling the reader to analyze predictions at specific timestamps~(\Cref{fig:results:multimodelcomponent}A), how quantities accumulate over time~(\Cref{fig:results:multimodelcomponent}B), and aggregates such as sums (totals) over a part or the entire experiment~(\Cref{fig:results:multimodelcomponent}C).

\subsection{Analysis and Configuration of Multi-Model Window Size} \label{sec:m3sa:multi-model:window-size}

\begin{figure}[t]
    \centering
    \includegraphics[width=0.95\linewidth]{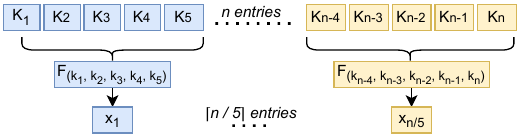}
    \vspace*{-0.15cm}
    \caption{Window size 5 applied on $n$ entries, using aggregation function $F$.}
    \label{fig:multimodel:windowing-concept}
    \vspace*{-0.45cm}
\end{figure}

\begin{figure}[t]
    \centering
    \includegraphics[width=0.95\linewidth]{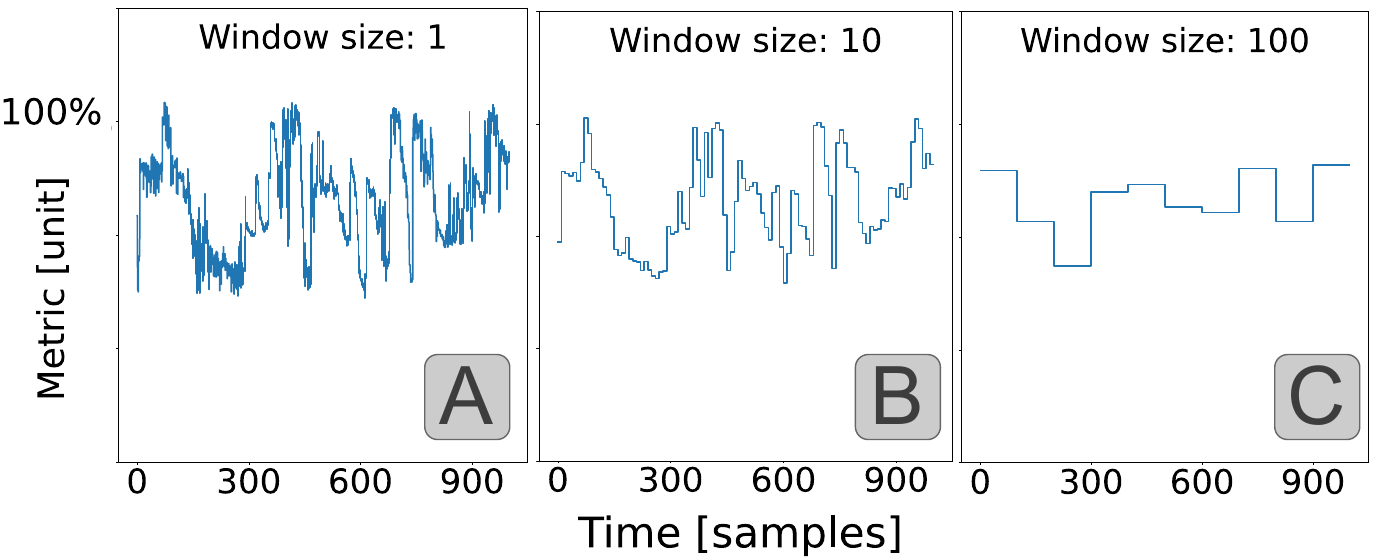}
    \vspace*{-0.15cm}
    \caption{Comparison of visual quality when windowing representative input data. Window sizes differ per sub-plot, in order, 1, 10, and 100. The horizontal axis represents time, in abstract units. The vertical axis represents the system response metric, from zero to maximum.}        
    \label{fig:f5-window-size-comparison}
\end{figure}

To address the tradeoff between
the detail of the visual comprehension~\ref{m3sa:fr1} and the performance~\ref{m3sa:nfr1} aspects of the multi-model, 
we equip the Multi-Model component with a windowing mechanism.  

We define a \textit{window} as an aggregation of chunks of data into a singular data entry, similar to a one-dimensional convolutional layer. 
\Cref{fig:multimodel:windowing-concept} illustrates how a window of size m (here, $m=5$) is applied to $n$ data entries. Each group of $m$ elements is fed into a configurable aggregation function $F$ (in this work, we use arithmetic mean), leading to a final compressed output array of size $\lceil n / m \rceil$.

We now explore the trade-off between visual quality and performance. 
We conduct experiments with actual outputs produced by the simulator, for input between about 2,000 and 200,000 samples long; with various window sizes between 1, corresponding to retaining all data, and 1000, a large value; and without M3SA-level optimizations such as eliminating redundancies. (When optimizations are enabled, as indicated by the results in \Cref{sec:experiments:exp1}, the performance of M3SA becomes compatible with~\ref{m3sa:nfr1}.)
\Cref{fig:f5-window-size-comparison} presents a representative selection of the results. 
As expected, the larger the window size, the sparser the filtered output and thus the coarser the visual detail. At a window size of 100, and beyond, the actual shape of the input is irretrievably lost. 
On the other side of the trade-off, the parsing and graphing sequence of our largest test-input ($n=201,600$), when $m=1$ requires, on a common, off-the-shelf, Apple M2 MacBook Pro, about 4.5\,minutes~(264.7\,s).
As the window size filters out data to persist and graph, the runtime of computation and graphing decreases as the window size increases, to just over 3\,minutes at $m=10$. From $m=10$, additional effects, e.g., the working set becoming increasingly large, reverse this trend, and the time needed when $m=1,000$ for the largest input already approaches the time taken without windowing~(238.3\,s).
Based on these results, we configure the Multi-Model with windowing-sizes of $m=1$~(\S\ref{sec:experiments:exp1}) and of $m=10~(\S\ref{sec:experiments:exp2}, \S\ref{sec:experiments:exp3})$ for this work.

\subsection{Design of Meta-Model component} \label{sec:m3sa:meta-model:design}\label{sec:m3sa:meta-model}

In this section, we detail the \textit{Meta-Model} component~(\Cref{fig:f3-m3sa-overview}, \circled{I}). Overall, the Meta-Model uses an automated process to combine predictions made with multiple models~\ref{m3sa:fr2} into a highly accurate result~\ref{m3sa:nfr2}. The strategy the Meta-Model follows is similar to bootstrap aggregating (bagging), a state-of-the-art technique adopted in the machine-learning community for improving accuracy and robustness, through prediction aggregation~\cite{DBLP:journals/ml/Breiman96b}.

The Meta-Model relies on the predictions made by multiple models. For performance reasons~\ref{m3sa:nfr1}, this process leverages processing results from the Multi-Model~(see~\Cref{sec:m3sa:multi-model:design}), thus avoiding redundant computation.  
Upon initialization, the core of the Meta-Model uses the Meta-predictor~(\circled{L}), which combines other models' (windowed) predictions \ref{m3sa:fr2}. After \circled{I} completes, the Meta-Model contains a one-dimensional set of predictions, prepared for output.

The output involves two independent processes, for usability~\ref{m3sa:fr1}, performance~\ref{m3sa:nfr1}, and portability~\ref{m3sa:fr3} reasons. First, the system compresses the predictions losslessly, into a Parquet data file, which the user can process outside of the M3SA system~(\circled{F}); we use Parquet due to the scalability, efficient storage, and cross-system compatibility (and thus portability) of this storage format~\cite{ApacheParquet}. Second, the Meta-Model is plotted~(\circled{E}) following user inputs~(\circled{A}).

We define \textit{robustness} as the confidence that predictions perfectly and consistently match the measured reality. Although the \textit{accuracy} of the Meta-Model is directly dependent on the accuracy of singular models~(\circled{I}) and on the aggregation function~(\circled{K}), the robustness of the Meta-Model accuracy should intuitively be higher than of any singular model~\ref{m3sa:nfr2}, because the Meta-Model can mitigate the errors of singular models when it combines the results across multiple models. This can be expressed and empirically measured for specific inputs, e.g., the MAPE metric introduced in~\Cref{sec:mesa:meta-model:mape}.

\begin{figure}[t]
    \centering
    \includegraphics[width=0.95\linewidth]{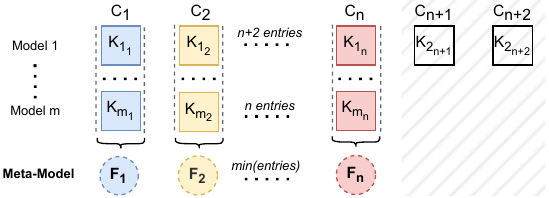}
    \vspace*{-0.15cm}
    \caption{The Meta-Model predictor, aggregating predictions of \textit{m} models vertically, for each time-step $C_1$ to $C_n$. Steps $C_{n+1}$ and $C_{n+2}$ are discarded.}
    \label{fig:meta-simulation-concept}
\end{figure}

To achieve high accuracy, the Meta-Model component uses the Meta-predictor~(\circled{L}). The Meta-predictor receives a set of predictions over time, one per input model, with time divided in equal chunks (steps). As \Cref{fig:meta-simulation-concept} illustrates, the Meta-predictor first aligns the steps across all models, for example, removing entries when too few models provide predictions, and aggregates the remainder. In the figure, the rows represent the predictions of each of the $m$ models, and the columns represent the aligned time-steps. Model 1 provides additional predictions, for time steps $C_{n+1}$ and $C_{n+2}$, which are discarded by the Meta-predictor because not enough other models also provide predictions for these steps.

Following alignment, the Meta-predictor applies an aggregation function. In the figure, the values stored in each column $C_{k}$ are aggregated for each time-step, using a (user-configurable) function $F_{k}$.
In \Cref{fig:meta-simulation-concept}, the aggregation function is applied vertically for each column, which is convenient also for performance. However, the concept also allows for diverse spatial combinations and further, more complex but model-order-free, approaches. 
We leave for future work the exploration of such aggregators, using mathematical, statistical, and machine-learning techniques.

For the aggregating function $F$, we explored for this work two computationally light, commonly used aggregation functions, (arithmetic) \textit{mean}, which is used in ecology~\cite{ecologyMultiModel} and meteorology~\cite{environmentMultiModel} but has high sensitivity to outliers, and \textit{median}, which alleviates the outlier sensitivity of the mean but is more computationally expensive.

The proposed Meta-Model design poses two challenges. Firstly, timestamp misalignments may occur due to non-deterministic simulation infrastructure. We address this challenge by selecting a datacenter simulator that establishes the same initial sampling timestamp for all models and the same sampling rate of the infrastructure. Secondly, singular models may output prediction datasets of different sizes (i.e., longer or shorter virtual infrastructure execution times), as we illustrate in \Cref{fig:meta-simulation-concept}, for models 1 and m; this difference may occur due to various reasons, such as scheduling algorithms that may prioritize tasks differently, or the simulation of potential infrastructure failures. To address this challenge, the meta-predictor uses only the minimum number of entries per model in the computation process.

This design of the Meta-Model already allows for reducing the error rates of peer-reviewed, state-of-the-art models (\S\ref{sec:experiments:exp1}). This design, yet still powerful, could be further explored by employing machine learning techniques in aggregating models, or by dynamically, on-the-fly, allocating weights to the singular models, or employing Multi-Criteria Decision Analysis (MCDA), which are widely used decision-making processes in other-scale sciences~\cite{mcda-huang2011multi, mcda-wang2009review, mcda-ishizaka2013multi}. Although this is beyond the purpose of this research, we envision future work which could analyze such techniques, trade-offs performance-accuracy, and advantages of more complex, yet more computationally heavy, aggregation techniques.

\subsection{Evaluation of Meta-Model Accuracy} \label{sec:mesa:meta-model:validation}\label{sec:mesa:meta-model:mape}

To quantify robustness~(see~\Cref{sec:m3sa:meta-model:design}) in M3SA~\ref{m3sa:nfr2}, we use the Mean Absolute Percentage Error~(MAPE), a widely used relative-error metric~\cite{oracle2014mape, DBLP:conf/wosp/NiewenhuisTIM24, moreno2013using-mape}. MAPE equally penalizes positive and negative errors and is calculated using \Cref{eq:mape}, where \textit{n} is the number of samples, \textit{R} is real-world data, \textit{S} is simulation data, \textit{i} is the sample index:

\vspace*{-0.5cm}
\begin{equation} \label{eq:mape}
    MAPE~[\%] = \frac{1}{n} \sum_{i=0}^{n} \left| \frac{R_{i} - S_{i}}{R_{i}} \right| \times 100
\end{equation}

Anticipating the need for further accuracy evaluation~\ref{m3sa:nfr2}, we design M3SA to be  extensible to other metrics such as Normalized Absolute Difference~(NAD)~\cite{DBLP:conf/wosp/NiewenhuisTIM24}, Root of the Mean Square Error~(RMSE)~\cite{rmse-chai2014root}, and Mean Average Error~(MAE)~\cite{DBLP:journals/jmlr/GunawardanaS09, DBLP:journals/tois/HerlockerKTR04}.

\begin{table*}
\small
\centering
\caption{Experiment configurations. (WT=workload trace, CT=CO2 trace, PM=performance metrics, SM=sustainability metrics, IM=individual models, MW=Multi-Model Window size, MF = Meta-Model aggregation function, SUO=system under observation.)}
\label{table:experiments-overall-design}
\vspace*{-.25cm}
\begin{tblr}{
  column{1} = {wd=0.01cm, leftsep=0pt, rightsep=0pt},
  column{3} = {c},
  column{6} = {c},
  column{10} = {r,},
  column{12} = {c},
  column{13} = {c},
  column{14} = {c},
  cell{1}{4} = {c=2}{c},
  cell{1}{7} = {c=5}{c},
  cell{1}{15} = {c},
  cell{2}{1} = {r=2}{},
  cell{2}{2} = {r=2}{},
  cell{2}{4} = {r=2}{},
  cell{2}{5} = {r=2}{},
  cell{2}{7} = {c=2}{c},
  cell{2}{9} = {c=3}{},
  cell{2}{13} = {r=2}{},
  cell{2}{14} = {r=2}{},
  cell{2}{15} = {r=2}{},
  cell{3}{9} = {r},
  cell{4}{9} = {r},
  cell{5}{9} = {r},
  cell{6}{9} = {r},
  hline{1,4,7} = {-}{},
  hline{2} = {4-5,7-11}{},
}
        &                                                            &  & Input    &     &  & Output    &         &                &    &        &  &          &  &     \\
& Focus                                                      &  & WT       & CT  &  & Metrics   &         & M3SA-Specifics &    &        &  & Failures &  & SUO \\
        &                                                            &  &          &     &  & PM        & SM      & IM             & MW & MF     &  &          &  &     \\
       & {Reproduce peer-reviewed\\experiment (§\ref{sec:experiments:exp1})}                      &  & WT1      & \ding{56}  &  & CPU hours & Wh      & 4              & 10 & median &  & \ding{56}       &  & S1  \\
       & {Analyze M3SA on fundamentally \\ different traces (§\ref{sec:experiments:exp2})} &  & WT2, WT3 & CT1 &  & CPU hours & gCO2    & 8              & 1  & mean     &  & \ding{52}       &  & S2  \\
       & {Adopt M3SA in CO2-aware, dynamic \\ workload migration §\ref{sec:experiments:exp3}}   &  & WT2      & CT2 &  & CPU hours & tCO2 & 16             & 1  & median &  & \ding{56}       &  & S3  
\end{tblr}
\vspace*{-.35cm}
\end{table*}

\begin{table}
\small
\centering
\caption{Systems Under Observation (SUO).}
\vspace*{-.15cm}
\label{table:systems-under-observation:suos}
\begin{tblr}{
  column{3} = {r},
  hline{1-2,5} = {-}{},
}
ID & Source  & Scale [\#hosts] & Resources per host            \\
S1 & SURF    & 277             & 128~GB RAM, 16 cores, 2.1~Ghz \\
S2 & Marconi & 150             & 196~GB RAM, 48 cores, 2.1~Ghz \\
S3 & Marconi & 2,982           & 196~GB RAM, 48 cores, 2.1~Ghz 
\end{tblr}
\vspace*{-.15cm}
\end{table}

\begin{table}
\small
\centering
\caption{Workload traces used in experiments. Name is source and collection year (e.g., SURF-22 = source SURF, year 2022). S = scientific, B = business-critical, SUO = system under observation, CH = CPU hours (millions), SR = sampling rate.}
\label{table:workload-traces}
\vspace*{-.25cm}
\begin{adjustbox}{width=\columnwidth,center}
\begin{tblr}{
  column{5} = {r},
  column{6} = {r},
  column{7} = {r},
  column{8} = {r},
  cell{1}{5} = {c},
  cell{1}{6} = {c},
  cell{1}{7} = {c},
  cell{1}{8} = {c},
  hline{1-2,5} = {-}{},
}
ID  & Name         & Type & SUO & Duration & Jobs  & CH        & SR  \\
WT1 & SURF-22      & S    & S1  & 7\,days    & 7,850 & 0.31   & 30\,s \\
WT2 & Marconi-22   & S    & S2  & 30\,days   & 8,316 & 4.74   & 20\,s \\
WT3 & Solvinity-13 & B    & S3  & 30\,days   & 50    & 0.13   & 30\,s 
\end{tblr}
\end{adjustbox}
\vspace*{-.15cm}
\end{table}
\begin{table}[t]
\centering
\caption{Carbon traces used in experiments. Name is source, location of collection, and collection year (e.g., ENTSOE-NL-22 = source ENTSOE, collected in the Netherlands, year 2022.)}
\label{table:carbon-traces}
\vspace*{-.15cm}
\begin{adjustbox}{width=\columnwidth,center}
\begin{tblr}{
  cell{2}{1} = {r},
  cell{2}{2} = {c},
  cell{2}{4} = {r},
  cell{2}{5} = {r},
  cell{3}{1} = {r},
  cell{3}{2} = {c},
  cell{3}{4} = {r},
  cell{3}{5} = {r},
  hline{1-2,4} = {-}{},
}
ID  & Name         & Location             & Duration & Sampling   \\
CT1 & ENTSOE-NL-22 & The Netherlands& 1\,year   & 900\,s \\
CT2 & ENTSOE-EU-23 & 29 European regions & 1\,year   & 900\,s 
\end{tblr}
\end{adjustbox}
\vspace*{-.15cm}
\end{table}

\section{Complex, Trace-based Experiments with M3SA} \label{sec:experiments}

In this section, we conduct diverse experiments, evaluating M3SA and simultaneously providing new insights into datacenter operation. 
To setup our experiments~(\Cref{sec:experiments:setup}), we first implement a prototype of M3SA, interface it with OpenDC, and provide the system open-source~\ref{m3sa:nfr2}. In \Cref{sec:experiments:exp1}, we reproduce a peer-reviewed experiment with M3SA capabilities and analyze M3SA against established requirements. In \Cref{sec:experiments:exp2}, we run an experiment using two fundamentally different traces, analyze the results, and evaluate the versatility of M3SA. Lastly, we evaluate M3SA on CO2-aware migration over energy-production patterns. We focus here on the main findings and further analyze the experiments in a technical report (Appendices A-D).

Overall, the experiments support 
four main findings (MFs):

\begin{enumerate}[label=(\textbf{MF\arabic*}),leftmargin=0pt,itemindent=3em]
    \item \label{experiments:mf1} The Multi-Model can increase the explainability of the simulation results, and help in identifying biases and errors of singular models; the Meta-Model can enhance prediction accuracy against singular models robustly. Specifically, the Meta-Model can reduce the error rate of the average singular simulation model by about 50\%~(MAPE decreases from 7.59\% to 3.81\%) with under 20\% computational overhead~(\Cref{sec:experiments:exp1}).
    
    \item \label{experiments:mf2} M3SA can simulate 1 (or 2) years of datacenter operation in 5 (7) minutes, of which under 0.5 (1) minutes are M3SA overhead~(\Cref{sec:experiments:exp1}).

    \item \label{experiments:mf3} Supporting various C-level decisions, M3SA can simulate business-critical and scientific workloads in scenarios including machine failures~(\Cref{sec:experiments:exp2}).
    
    \item \label{experiments:mf4} Adding to MF3, M3SA can help in CO2-aware workload migration and scheduling processes, with decisions that can help reduce CO2 emissions by 97.5\% compared to the average location, and by 11\% compared to the best but single location~(\Cref{sec:experiments:exp3}).

\end{enumerate}

\subsection{Experiment Setup} \label{sec:experiments:setup}

We summarize our experimental design in \Cref{table:experiments-overall-design}, and detail the infrastructure, workloads, and carbon traces in \Cref{table:systems-under-observation:suos}, \Cref{table:workload-traces}, and \Cref{table:carbon-traces}, respectively. We detail the most important aspects:

\textit{Workload traces~(all public):} (WT1) SURF-22 is a scientific workload trace used in peer-reviewed experiments~\cite{DBLP:conf/wosp/NiewenhuisTIM24}. SURF-22 was traced in one of the largest HPC facilities at SURF, the Netherlands. It traces scientific jobs run in production, batch,
with an average duration of 39.52\,CPU-hours, and also detailed energy use.

(WT2) Marconi-22 is one of the most detailed open-source workload traces currently available. It traces scientific jobs, multiple operational layers, and the electrical network of Marconi HPC, a powerful supercomputing facility hosted by CINECA in Italy~\cite{DBLP:journal/nature/BorghesiSDMBGMCGCBBB23}.

(WT3) Solvinity-13 is a business-critical workload hosted in a Solvinity datacenter~\cite{site:solvinity}, formerly BitBrains, a mid-sized cloud service provider in the Netherlands. Solvinity-13 has been used in highly cited, peer-reviewed publications on business-critical workloads~\cite{DBLP:conf/ccgrid/ShenBI15}. Solvinity-13 traces long-running jobs, with an average of 2,722\,CPU-hours per job.

\textit{Carbon traces (public):} ENTSO-E Transparency Platform~\cite{site:entso-e} consists of 40 European Transmission System Operators (TSOs), providing open source datasets about energy and sustainability metrics since 2015. In this work, we use traces from 29 countries from the ENTSO-E datasets.

\textit{Failure model:} 
The failure model causes infrastructure failures at arbitrary times, with downtime periods following Ldns04 traces~\cite{DBLP:conf/ccgrid/KondoJIE10}, which follow an exponential distribution with known parameters~\cite{DBLP:conf/ccgrid/KondoJIE10}. We assume no checkpointing system, so each job is re-run from the beginning after failure.

\textit{Power (draw) models:} We only model power draw due to CPU usage in this work. For the Multi-Model (\S\ref{sec:m3sa:multi-model:design}), we equip the simulator with a library of seven commonly used power-draw models~\cite{DBLP:conf/iccS/SilvaOCTDS19, DBLP:journals/spe/CalheirosRBRB11, DBLP:conf/ccgrid/MastenbroekAJLB21, DBLP:conf/isca/FanWB07}, each with a simple analytical formula that links power draw (output) to idle and maximum power draw, and to current utilization; the models include square root, linear, square, cubic, MSE, asymptotic, and asymptotic with cubic attenuation due to DVFS ($DVFS(u) = P_{\text{idle}} + \frac{(P_{\text{max}} - P_{\text{idle}})}{2} \times( 1 + u^3 - e^{-u^3/\alpha})$, where $u$ = CPU utilization, $P_{idle}$ and $P_{max}$ are the power used in idle and full capacity states, and $\alpha$ is the utilization fraction at which the host becomes asymptotic). In experiments 1 (\S \ref{sec:experiments:exp1}), 2 (\S \ref{sec:experiments:exp2}), and 3 (\S \ref{sec:experiments:exp3}), we select 4, 8, and 16 combinations of models and specific realistic parameters, respectively. For example, in experiment 2, we use each of the seven power-draw models, and configure only the last with two different configurations, $\alpha=0.85$ and $\alpha=1.90$, and $P_{idle}=32$, $P_{max}=180$. We detail each power (draw) model in the technical report, available in \Cref{sec:appendix:a}.

\subsection{Peer-reviewed experiment reproduced with M3SA}\label{sec:experiments:exp1}

\begin{figure}[t]
    \centering
    \includegraphics[width=0.95\linewidth]{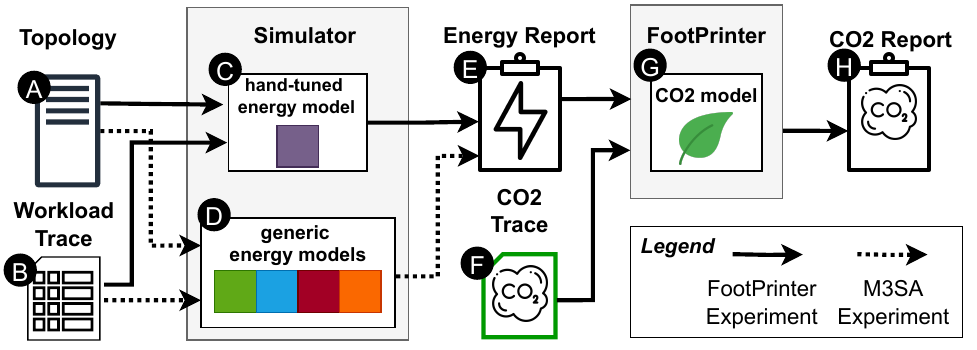}
    \caption{Overview of experiment 1, adapted and re-run from Footprinter~\cite{DBLP:conf/wosp/NiewenhuisTIM24}, enhanced with M3SA capabilities.}
    \label{fig:exp1-overview}
    \vspace*{-0.5cm}
\end{figure}

Niewenhuis et al. propose FootPrinter~\cite{DBLP:conf/wosp/NiewenhuisTIM24}, a tool for predicting the CO2 footprint of datacenters. \Cref{fig:exp1-overview} depicts the experiment we reproduce from~\cite{DBLP:conf/wosp/NiewenhuisTIM24}, leveraging M3SA capabilities to achieve higher accuracy and more detailed explanations; \Cref{sec:experiments:setup} and \Cref{table:experiments-overall-design} summarize the design of this experiment. 
Illustrative for the differences between approaches, whereas Footprinter employs a single, hand-tuned energy model~(\circled{C}), M3SA uses multiple, generic energy models~(\circled{D}). The other components of the experiment, \circled{A}, \circled{B}, and \circled{E} to \circled{H}, remain unchanged, enabling a direct comparison between approaches.
Our results strongly support MF1.

\begin{figure}[t]
    \centering
    \includegraphics[width=0.85\linewidth]{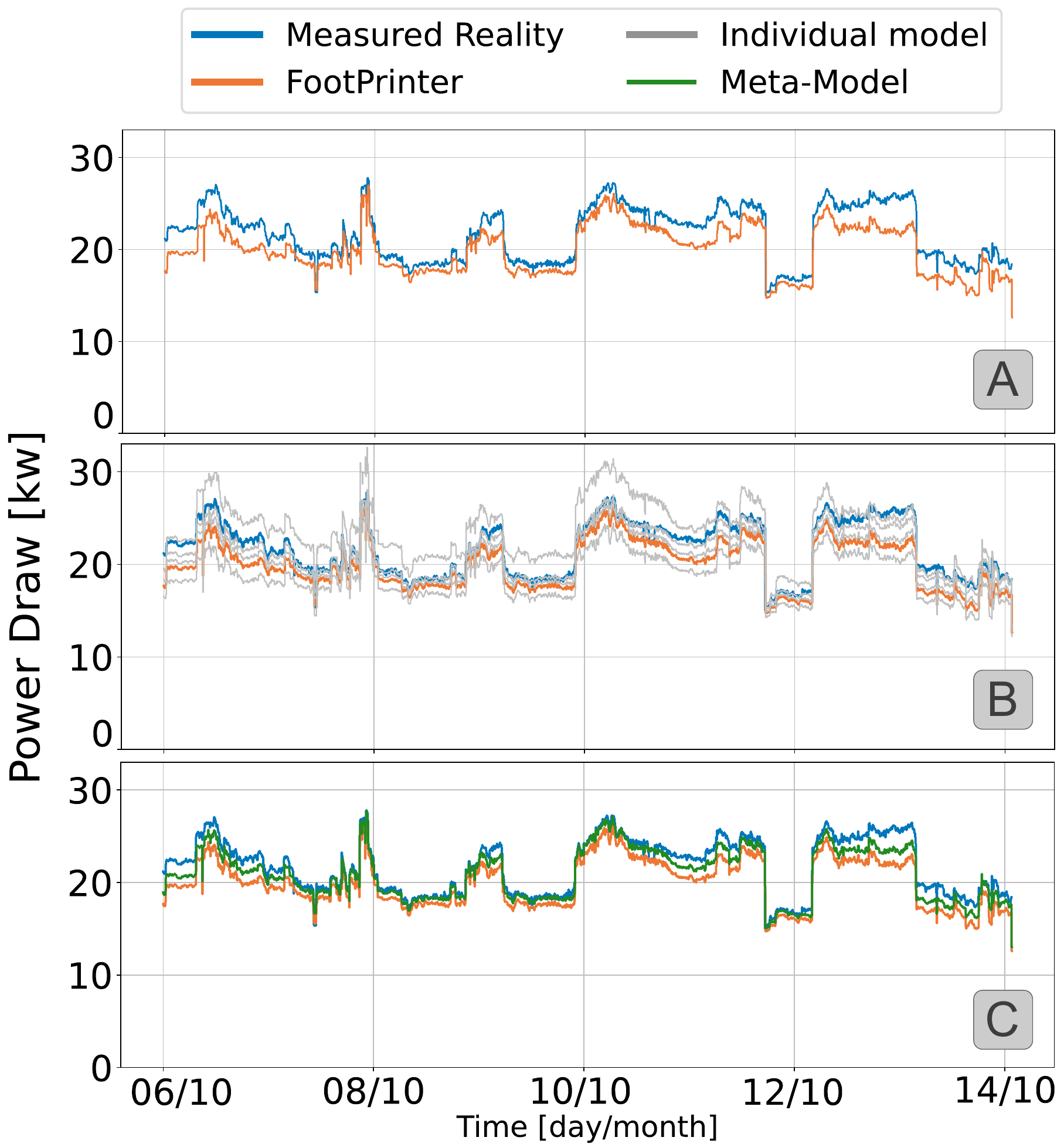}
    \caption{Simulation results vs. measured reality: (A) FootPrinter simulator, peer-reviewed experiment~\cite{DBLP:conf/wosp/NiewenhuisTIM24}; (B) Multi-Model with four singular models vs. FootPrinter; (C) Meta-Model derived from the Multi-Model at point B.}
    \vspace*{-0.5cm}
    \label{fig:exp1:A-B-C:results}
\end{figure}

\Cref{fig:exp1:A-B-C:results} shows the power draw of the S1 system over one week. \Cref{fig:exp1:A-B-C:results}A shows the results of reproducing the experiment from FootPrinter~\cite{DBLP:conf/wosp/NiewenhuisTIM24}, compared against measured reality. 
Illustrative for a hand-tuned model, the FootPrinter curve approaches in many places the measured reality. As a trade-off, changing the model for a new setting would be complex and error-prone, requiring possibly a similar amount of work to the development of the initial model. (We explore this ability for M3SA, in \Cref{sec:experiments:exp2}.)

\textit{Better explainability}: \Cref{fig:exp1:A-B-C:results}B additionally shows four singular models, whose results were automatically produced through the Multi-Model component. 
It seems appealing to select the most accurate model, from this sub-plot or peer-reviewed publications, by comparing against the ground truth provided by the measured reality.
However, in real-world situations, and increasingly when opaque clouds provide resources, the analyst is often unaware of the ground-truth. 
The results in \Cref{fig:exp1:A-B-C:results}B show that three of the singular models match the measured reality, while a fourth model constantly overestimates. Using the Multi-Model approach, the analyst can identify singular model biases by contrasting the predictions at each moment in time, thus increasing the overall prediction explainability\footnote{Resolving the discrepancies between models, and identifying a single best model, requires capabilities beyond the scope of this work.} over singular-model approaches, where inaccuracies may not even be observable.

\textit{``All models are wrong, but some are useful`` (George Box, 1976).} Simulating with a robustly accurate model is a critical yet non-trivial challenge in ICT. \Cref{fig:exp1:A-B-C:results}C shows the predictions of a Meta-Model, automatically derived from the Multi-Model presented in \Cref{fig:exp1:A-B-C:results}B, and compared against the predictions of the FootPrinter model and the measured reality. We quantify accuracy against measured reality and obtain a MAPE of 7.59\% for the average singular model, and a MAPE of 3.81\% ($\approx$50\% better) for the Meta-Model. The improvement in error rate is due to the \textit{median} aggregation function, which effectively filters extremes, reduces biases, and improves the robustness of the prediction. 

Compared to the state-of-the-art FootPrinter's MAPE of 3.15\%, the Meta-Model has a 
higher error rate. However, FootPrinter underlies a manual simulation model, explicitly trained for a single trace (here, SURF-22).
In contrast, M3SA enables the versatile, automatic re-use of generic models.

\begin{figure}
    \centering
    \includegraphics[width=0.95\linewidth]{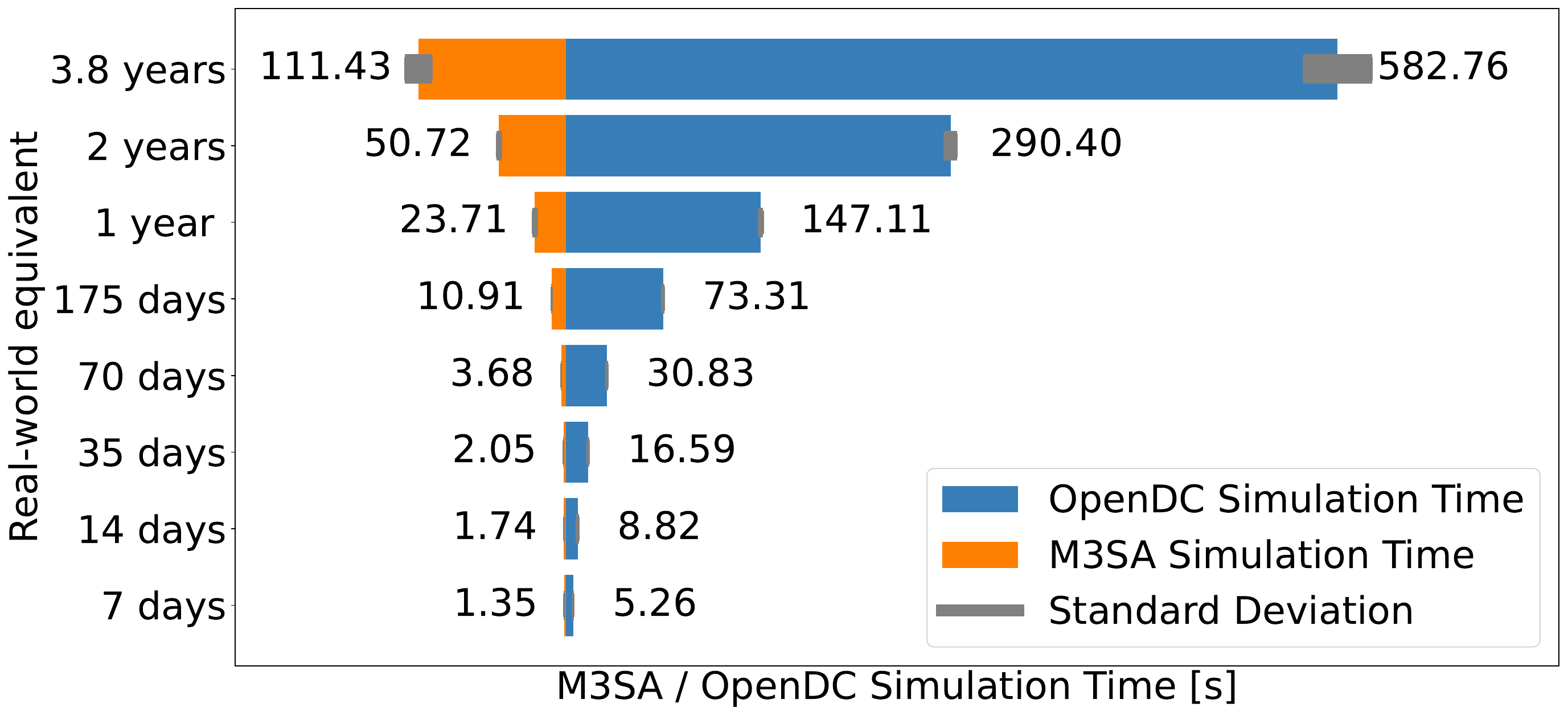}
    \vspace*{-0.1cm}
    \caption{M3SA performance overhead compared to OpenDC simulation runtime.
    The plot, which takes its structure from Tukey's population pyramid, plots on the horizontal axis the system response: (left) M3SA's overhead, (right) the OpenDC simulation time. On the vertical axis, discretely, is the duration of the workload; higher values cause higher runtimes. 
    }
    \vspace*{-0.55cm}
    \label{fig:exp1-nfr1-performance}
\end{figure}

\textit{M3SA runtime overhead under realistic conditions:}
Throughout this experiment, M3SA operated on 20,170 samples, or about 7 days of operation with SURF monitoring rate of 30 seconds~\cite{DBLP:conf/ccgrid/MastenbroekAJLB21}. To assess the runtime overhead of M3SA, we further create datasets of various sizes, equivalent to different datacenter operational times, including 2\,years~\ref{m3sa:nfr3} and up to nearly 4\,years, by modifying the monitoring granularity of the simulator. 

\Cref{fig:exp1-nfr1-performance} summarizes the results of this performance-analysis experiment. 
The M3SA overhead includes synthesizing a Multi-Model, computing the Meta-Model, and outputting the results. 
We run each experiment 10 times on a standard MacBook M2 Pro and represent the standard deviation in gray (full analysis in~\cite{m3sa-technical-report}, Appendix B).
On average, M3SA adds less than 20\% time overhead to the overall simulation process, even for massive-scale datasets, and can simulate years of datacenter operation within minutes~\ref{m3sa:nfr1}.

\subsection{Evaluating M3SA on fundamentally different traces}\label{sec:experiments:exp2}

\begin{figure}[t]
    \centering
    \includegraphics[width=0.95\linewidth]{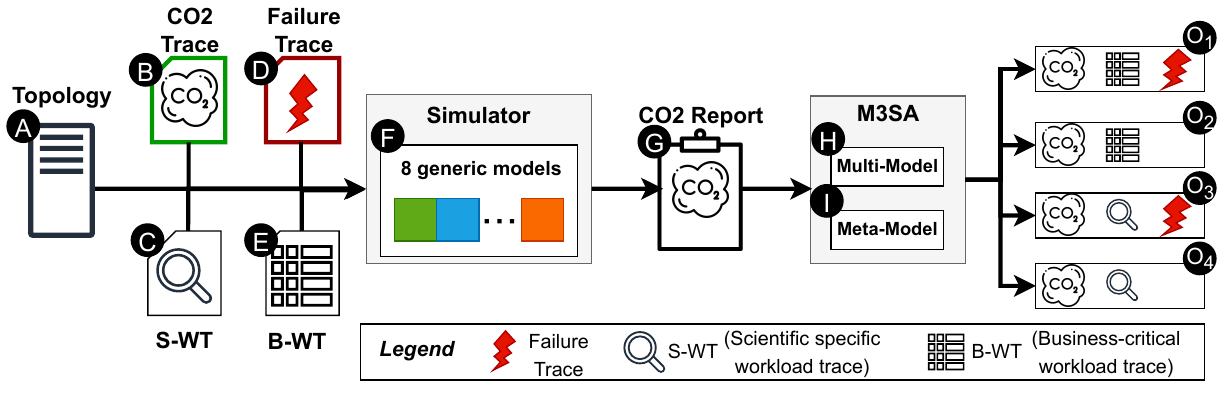}
    \caption{Experiment 2 overview. Evaluating M3SA on business-critical and scientific-specific workloads, with operational phenomena.}
    \label{fig:exp2-overview}
    \vspace*{-0.5cm}
\end{figure}

Support for NFR2 also introduces the question of which models can be considered. 
The ability to cope with the diversity of possible workloads is already understood as essential for datacenter simulators~\cite{DBLP:conf/usenix/AmvrosiadisPGGB18}.
Similarly, failures are known to change how specific workloads are run~\cite{DBLP:conf/sigcomm/GillJN11, DBLP:conf/usenix/AmvrosiadisPGGB18}.
We showcase in this section, through an experiment, how M3SA copes with fundamentally different traces and with machine-failures.
\Cref{sec:experiments:setup} and \Cref{table:experiments-overall-design} summarize the design of this experiment. \Cref{fig:exp2-overview} depicts the experiment design including two kinds of workloads to be run independently~(\circled{C} and \circled{E}), and the option to include failures~(\circled{D}); these lead to four output combinations, \circled{$O_1$} to \circled{$O_4$}.

\begin{figure}[t]
    \centering
    \includegraphics[width=0.95\linewidth]{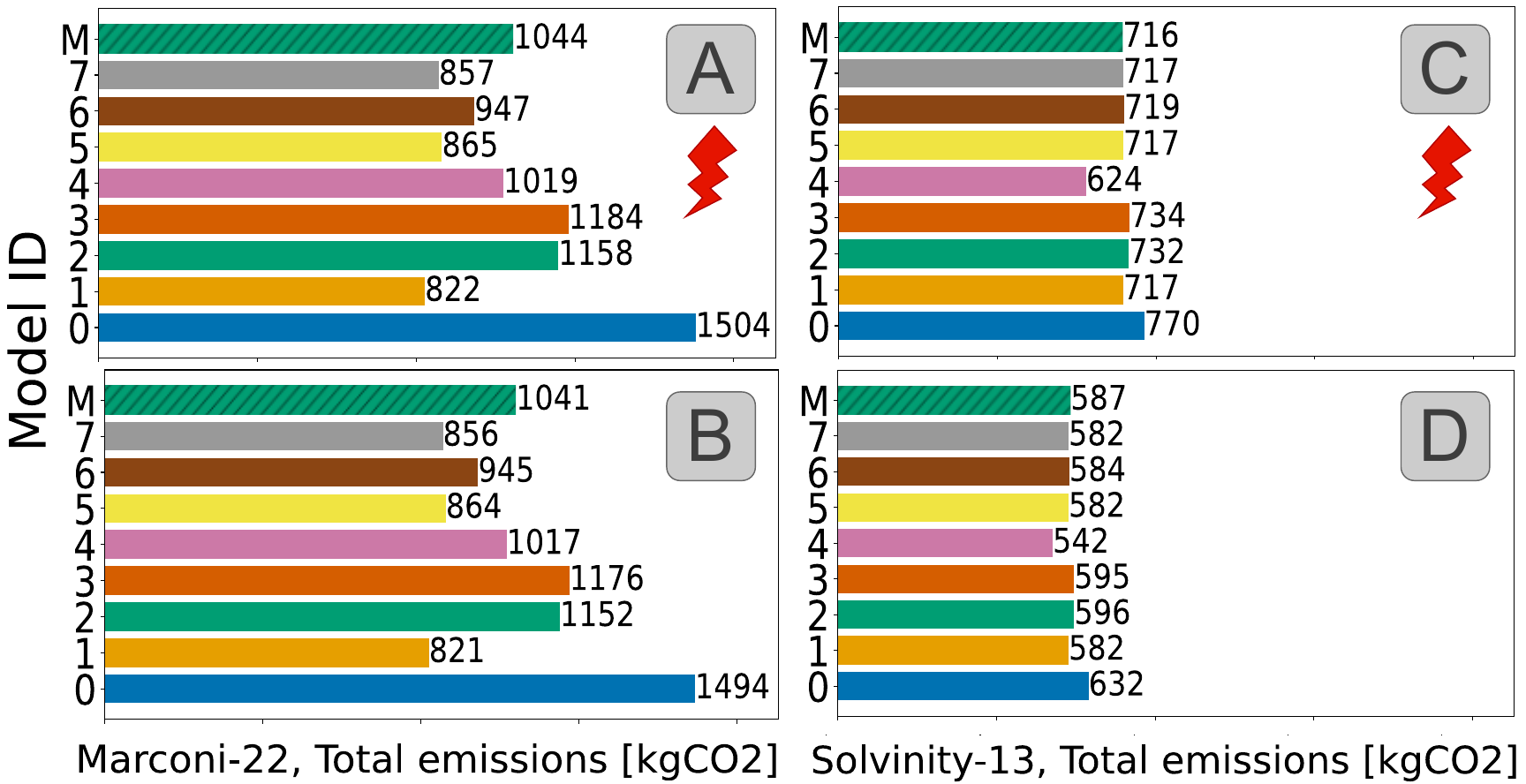}
    \caption{Total emissions generated by a workload trace running in S2. Left column, Marconi-22 workload: (A) S2 experiences Ldns04-based failures, (B) S2 without failures. Right column, Solvinity-13 workload: (C) S2 experiences Ldns04-based failures, (D) S2 without failures.
    Each plot shows the total CO2 emissions predicted by the M3SA Multi-Model using eight generic energy-models (labeled 0-7) and by the Meta-Model (M).}
    \label{fig:exp2-failure-comparison-2-traces}
    \vspace*{-0.35cm}
\end{figure}

Overall, besides the ability of M3SA to provide answers for these complex conditions~\ref{experiments:mf3}, the results depicted in \Cref{fig:exp2-failure-comparison-2-traces} indicate
that the presence of
failures can add from near-zero (0.28\%) to significant (21.9\%) CO2 emissions compared to running the workload without system failures, so sustainability studies such as~\cite{market:IDC24AI} should explicitly include such analysis. Consequently, C-level stakeholders should not make decisions absent such analysis. 

We run two fundamentally different traces. Not only do they come from different domains, scientific vs. business-critical, but also, whereas Marconi includes many job arrivals that fit diurnal and day-of-week patterns~\cite{DBLP:journal/nature/BorghesiSDMBGMCGCBBB23}, Solvinity-13 has much longer jobs~(see~\Cref{sec:experiments:setup}) and forms a stable workload that is largely time-insensitive~\cite{DBLP:conf/ccgrid/ShenBI15}.

\Cref{fig:exp2-failure-comparison-2-traces} shows the environmental impact of the two different workloads, when failures are not or also considered.
The results support that M3SA can help identify (overly) biased models, even when considering only the aggregate results. In this experiment, we observe that model 0 (square root, introduced in \S\ref{sec:experiments:setup}, $sqrt(u) = P_{\text{idle}} + (P_{\text{max}} - P_{\text{idle}}) \times \sqrt{u}$) constantly overestimates, predicting up to 54\% more than the average of the other models (value computed from Figure \ref{fig:exp2-failure-comparison-2-traces}, subplot A, where model 0 predicts 1,504~kgCO2, $\approx$54\% more than the average of 979~kgCO2 of models 1-7, value computed as (1,504 - 979) / 979 $\approx$ 54\%).
This major overestimation, albeit clear in a Multi-Model simulation, would not be observable in a single-model simulation; here, the Meta-Model approach helps.

The reporting features of M3SA allow for tailored evaluation and analysis. 
For example, when running Marconi-22,
the presence of failures leads to only a 0.28\% increase in CO2 emissions~(\Cref{fig:exp2-failure-comparison-2-traces}, sub-plots A and B, value computed as the difference between the Meta-Model results, labeled M in each sub-plot, normalized by the smallest of the Meta-Model results, i.e., $(1,044-1,041)/1,041$). 
In contrast, when running Solvinity-13, which has a much longer job duration, failures add 21.9\% CO2 emissions, equivalent to 129~kg of CO2~(\Cref{fig:exp2-failure-comparison-2-traces}C and D). Scaling linearly to the scale of Marconi HPC, consisting of 2,982 nodes, operational phenomena would add 2.56 tons of CO2. 
M3SA can aid C-levels in predicting such events, thus reducing the risk of unexpected CO2 emissions and can help operators cope with increasingly tight climate-related regulation.

\subsection{Evaluating M3SA on CO2-aware workload migration}\label{sec:experiments:exp3}

ICT infrastructure accounts for 3\% of global greenhouse gases, on par with the aviation industry~\cite{DBLP:conf/asplos/Eeckhout24}. 
To address this issue, C-levels adopt CO2 reporting and instruct their organizations to adopt CO2-aware scheduling and migration policies while still respecting Service Level Agreements~\cite{DBLP:conf/igsc/SouzaJCBLSAIS23}; 
in the upcoming years, regulation could require even more detailed analysis and related actions. 
Overall, CO2-aware spatial shifting, at the global level, can reduce CO2 emissions by up to 96\%, assuming infinite datacenter capacity, and up to 51\%, assuming finite capacity~\cite{DBLP:conf/eurosys/SukprasertSBIS24}. Still, how the frequency of workload migration impacts CO2, or how workload migration inside Europe impacts CO2, remains largely unexplored.

\begin{figure}[t]
    \centering
    \includegraphics[width=0.95\linewidth]{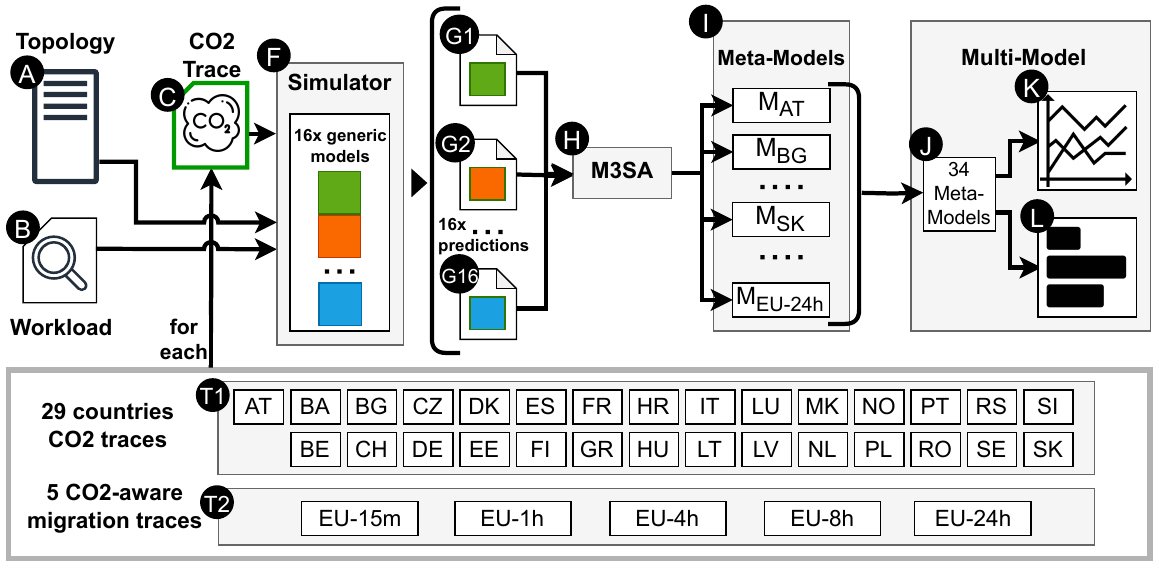}
    \vspace*{-0.15cm}
    \caption{Overview of experiment 3, assessing CO2 footprint of a workload run in 29 countries, and run using 5 CO2-aware migrations intervals.} 
    \label{fig:exp3-overview}
    \vspace*{-0.65cm}
\end{figure}

In this section, we analyze how M3SA copes with CO2-aware workload scheduling. \Cref{fig:exp3-overview} depicts the experiment including CO2 traces from 29 European countries~\circled{T1} and 5~migration traces at different granularities, where the location with the lowest CO2 impact is chosen at given intervals~\circled{T2}. As summarized by \Cref{table:experiments-overall-design}, we equip M3SA with the same 16 singular energy models and ICT infrastructure in each experiment. To enhance prediction robustness and alleviate per-model biases, we synthesize a Meta-Model for each location. Then, we run a greedy CO2-aware migration algorithm and obtain 5 CO2-aware migration traces~(\circled{T2}).

\begin{figure}[t]
    \centering
    \includegraphics[width=0.95\linewidth]{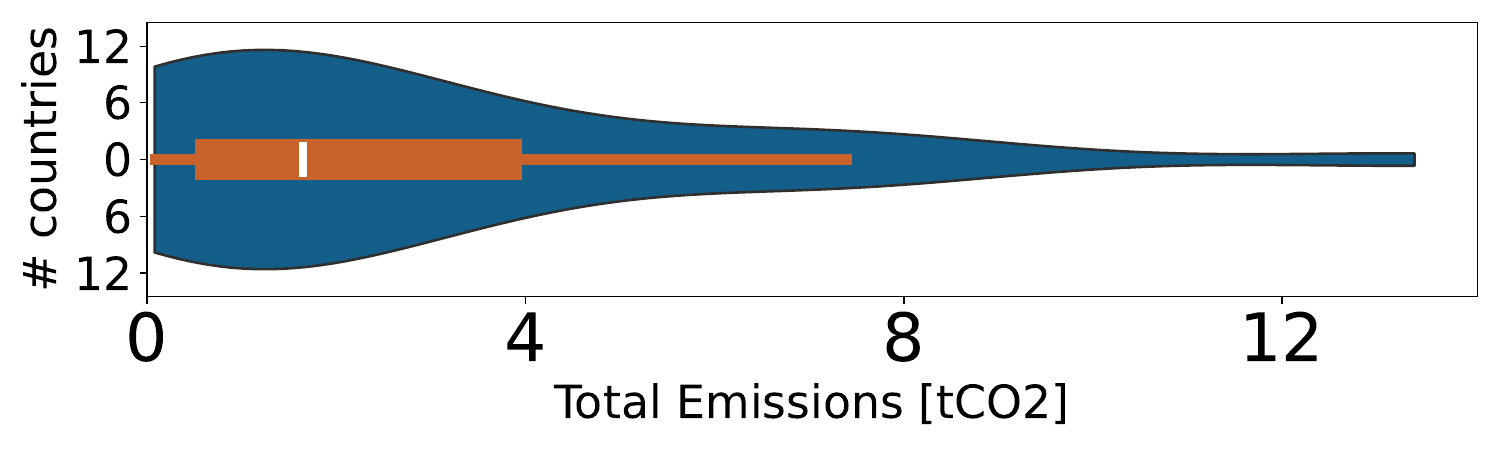}
    \vspace{-0.3cm}
    \caption{Total CO2 emitted when running the Marconi-22 workload on S3, assessed for 29 European countries, as of June 2023. The orange box and extended whiskers represent the typical box-and-whiskers plot, overimposed on a violin plot; we predict with one Meta-Model per country.}
    \vspace{-0.3cm}
    \label{fig:exp3-total-emissions-violin}
\end{figure}

Overall, our results show that CO2-aware workload migration and scheduling processes can help reduce CO2 emissions significantly~\ref{experiments:mf4} and should be considered more extensively by operators during both capacity planning and operational phases. \Cref{fig:exp3-total-emissions-violin} shows the distribution of CO2 emissions, for each considered location, without migration. We identify considerable correlation between the location of the datacenter and the amount of CO2 emissions; in this experiment, the improvement factor is significant, and up to 160x~(e.g., Germany, 13~tons~CO2 vs. Switzerland, ~0.081~tons~CO2).

We design and prototype a greedy CO2-aware migration algorithm (Appendix~\cite{m3sa-technical-report}), which migrates the workload to the greedy-best CO2 location, assuming no migration costs, instant migration, and sufficient infrastructure capacity; we select migration intervals of 15 minutes, 1 hour, 4 hours, 8 hours, and 24 hours~(\circled{T2}).

We analyze the number of migrations for each month of 2023 and observe that June has the most overall migrations, while January has the least. We run Marconi-22 workload, spanning over 30 days, in June 2023 and observe the CO2 footprint distribution ranging from less than 100~kg~CO2 (e.g., Sweden, Switzerland) up to 13~tons~CO2 (e.g., Germany), with an average of 2.92~tons~CO2. The differences, some in orders of magnitude, reflect the CO2 intensity of the energy sources; for example, in Switzerland, less than 1\% of electricity was obtained from brown sources~\cite{site:switzerland_2023_co2}, compared to 37\% in Germany~\cite{site:germany_2023_co2}.

\begin{figure}[t]
    \centering
    \includegraphics[width=0.95\linewidth]{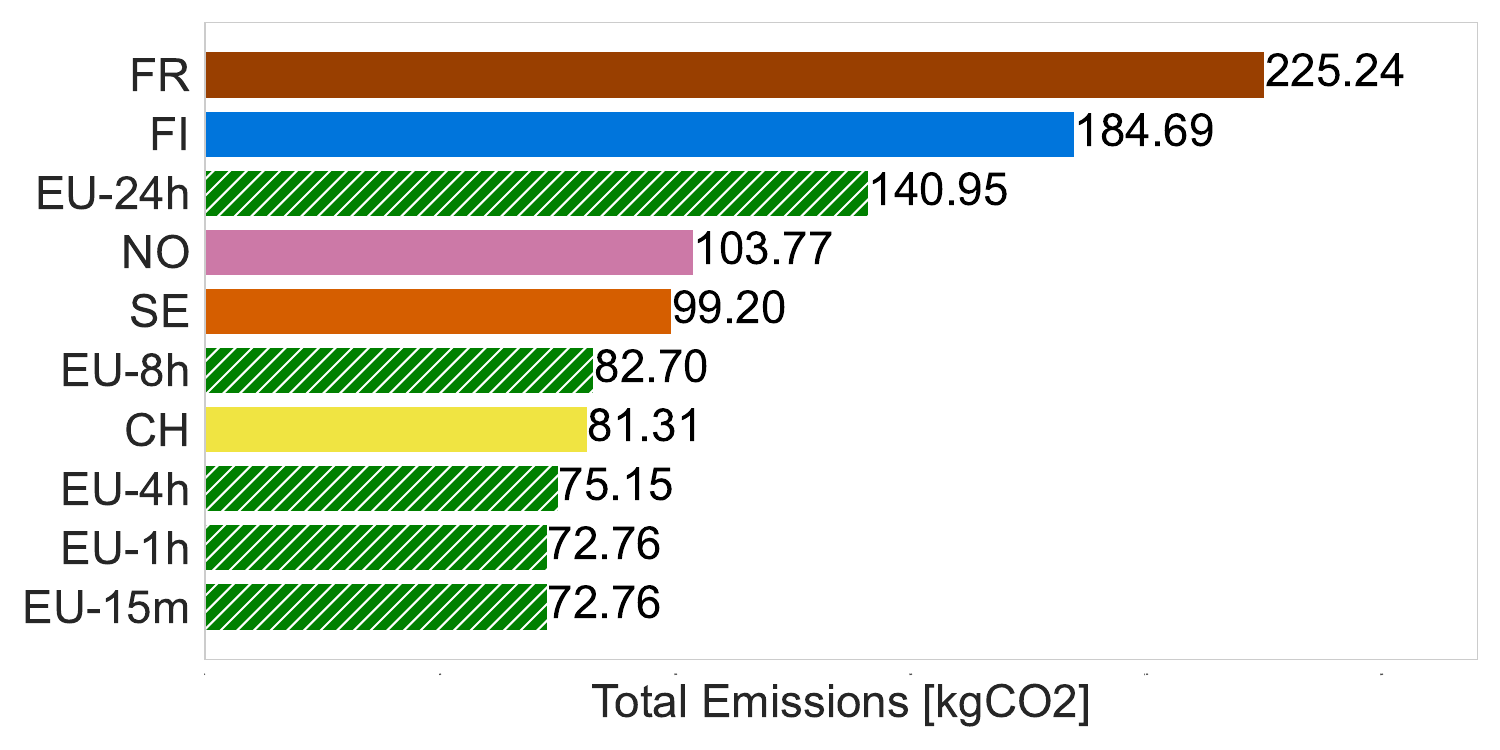}
    \vspace{-0.3cm}
    \caption{Configurations with ten-lowest cumulated CO2 -emissions. Migration within EU is considered with different temporal granularities, from 15\,minutes to 24\,hours.}
    \vspace{-0.5cm}
    \label{fig:exp3-top-10-locations}
\end{figure}

We identify migration intervals of 15 minutes and 1 hour as emitting 11\% less CO2 than running the workload integrally in the lowest CO2 location, and 97.5\% fewer emissions, equivalent to 2.9 tons than running the workload integrally in an average location; \Cref{fig:exp3-top-10-locations} shows top-10 CO2 locations (vertical axis) and cumulated CO2 emissions (horizontal axis). Still, while 24-hour migrations are more sustainable than 26 out of 29 locations, daily migrations can result in 73\% more emissions than running the workload in Switzerland, the lowest CO2 location.

The reporting features of M3SA allow for tailored CO2-aware migration and analysis and aid the migration algorithm in selecting the locations by CO2 footprint. This experiment unveils a tradeoff between migration granularity and environmental impact, unveiling new research areas in which M3SA can help identify, compare, and evaluate CO2-aware migrators and schedulers. Such tools would assist datacenter operators in making better-informed decisions about workload scheduling and migration, significantly improving operational footprint.

\section{Related Work} \label{sec:related-work}

Simulating using multiple models is a novelty in computer systems; however, as introduced in \Cref{sec:introduction}, other scientific fields already employ multi-model simulations, such as machine learning (ensemble learning, bagging~\cite{DBLP:journals/ml/Breiman96b}), statistical modelling, virology~\cite{covid19ForecastHubUS}, weather and climate simulation~\cite{schunk2016space-multimodel-weather}, or ecology~\cite{ecologyMultiModel}. 
Furthermore, existing ICT simulators (e.g., OpenDC~\cite{DBLP:conf/ccgrid/MastenbroekAJLB21}, CloudSim~\cite{DBLP:journals/spe/CalheirosRBRB11}) or general simulation frameworks (e.g., Mosaik \cite{app9050923}) enable co-simulation by combining and coordinating individual models or tools, yet each for a different purpose (e.g., a model for CPU utilization, a model for CO2 emissions, and a simulation tool for datacenter thermal system). 
In ICT, existing simulators currently rely on singular model predictions~\cite{DBLP:conf/ccgrid/MastenbroekAJLB21, DBLP:journals/spe/CalheirosRBRB11, DBLP:journals/spe/HewageIRB24, DBLP:conf/ccgrid/Casanova01, DBLP:journals/fgcs/McDonaldDWSC24}. 

Closest to our work, we identify COVID-19 Forecast Hub~\cite{covid19ForecastHubUS}, used to predict the COVID-19 hospitalizations, and providing 93 models, out of which users can select and leverage in a unified visualization. This is equivalent to the \textit{Multi-Model} component of M3SA but for a different field.  In weather and climate simulation, Myhre et al.~\cite{environmentMultiModel} investigate harmful emissions using seven models, aggregated (averaged) into a ``Mean Model''; this is equivalent to the \textit{Meta-Model} component of M3SA.

Datacenter simulators are useful tools for the community and are widely used in predicting ICT infrastructure. However, prior to this work, no simulator provides multi- and meta-simulation capabilities but relies on singular-model predictions~\cite{DBLP:conf/ccgrid/MastenbroekAJLB21, DBLP:journals/spe/HewageIRB24, DBLP:journals/spe/CalheirosRBRB11, DBLP:conf/ccgrid/Casanova01, DBLP:journals/fgcs/McDonaldDWSC24}.

\section{Conclusion and Future Work} \label{sec:conclusion}
Understanding the performance and climate impact of existing and, more importantly, upcoming datacenters is essential to our society and economy. 
Although datacenter operators already use simulators to predict infrastructure capabilities, current simulators often rely on singular models, which are insufficient for reliable predictions under diverse scenarios.
 
In this work, we have designed, implemented, and evaluated M3SA, a framework for multi- and meta-model simulation and analysis of datacenters. 
We evaluated M3SA using real-world workload and CO2 emission traces, demonstrating M3SA's ability to predict in complex scenarios, with enhanced explainability and robustness over approaches that only use singular models (even hand-tuned). 
Such capabilities can aid analysts and C-level decision-makers to better reason about real-world scenarios and make informed decisions. Results show that M3SA's can simulate years of datacenter operation in minutes, with a 50\% lower error rate than singular models, while enabling detailed explanation and without significant overheads. We also showcase how M3SA can simulate system failures and contribute to CO2-aware workload migration and scheduling, reducing CO2 emissions significantly when considering datacenters' geographical location.

We have released M3SA as an open-sourced system that can be applied as a top layer to datacenter simulators and tested its operation with the commonly used open-source simulator OpenDC. 
As explained in \Cref{sec:m3sa:meta-model:design}, we envision future research in exploring MCDA, and various methods of leveraging a Meta-Model, such as dynamically allocating weights to predictions of singular models.
We also envision future research in employing AI/ML techniques in the Meta-Model aggregation function 
and are currently embedding M3SA into a simulation-based \textit{digital twin}~\cite{NAP26894}, as part of a major infrastructure project with over 75 partner institutions.

\begin{acks}
This work was partially supported by EU MSCA CloudStars (g.a. 101086248) and EU Horizon Graph Massivizer (g.a. 101093202). This research is partly supported by a National Growth Fund through the Dutch 6G flagship project “Future Network Services.”
\end{acks}

\bibliography{refs}

\clearpage
\appendix
\section*{Technical Report}\label{sec:technical-report}
In Appendices A - C, we present a technical report, complementary to the experiments run in \Cref{sec:experiments}.

\appendix

\section{Experiment Overview}\label{sec:appendix:a}

\begin{table*}[t]
\centering
\caption{Formulas of the power models used in this work. E=experiment (e.g., E1=experiment 1). $P_{idle}$, $P_{max}$ = the power used in idle and full capacity states, $u$ = CPU utilization, $e$ = Euler's Number, $\alpha$ = utilization fraction at which the host becomes asymptotic, $r$ = calibration parameter.}
\label{table:model-formulas}

\begin{tblr}{
  colspec={l l X[l] l ccc},
  hline{1-2,9} = {-}{},
}
ID  & Name             & Formula                                                                  & Source                                                                            & E1 & E2 & E3 \\

EQ1 & Sqrt             & $\displaystyle 
                          P(u)=P_{\text{idle}}+(P_{\text{max}}-P_{\text{idle}})\sqrt{u}$          & \cite{DBLP:conf/iccS/SilvaOCTDS19,DBLP:journals/spe/CalheirosRBRB11,DBLP:conf/ccgrid/MastenbroekAJLB21} & \ding{52} & \ding{52} & \ding{52} \\

EQ2 & Linear           & $\displaystyle 
                          P(u)=P_{\text{idle}}+(P_{\text{max}}-P_{\text{idle}})u$                 & \cite{DBLP:conf/iccS/SilvaOCTDS19,DBLP:journals/spe/CalheirosRBRB11,DBLP:conf/ccgrid/MastenbroekAJLB21} & \ding{56} & \ding{52} & \ding{52} \\

EQ3 & Square           & $\displaystyle 
                          P(u)=P_{\text{idle}}+(P_{\text{max}}-P_{\text{idle}})u^{2}$             & \cite{DBLP:conf/iccS/SilvaOCTDS19,DBLP:journals/spe/CalheirosRBRB11,DBLP:conf/ccgrid/MastenbroekAJLB21} & \ding{56} & \ding{52} & \ding{52} \\

EQ4 & Cubic            & $\displaystyle 
                          P(u)=P_{\text{idle}}+(P_{\text{max}}-P_{\text{idle}})u^{3}$             & \cite{DBLP:conf/iccS/SilvaOCTDS19,DBLP:journals/spe/CalheirosRBRB11,DBLP:conf/ccgrid/MastenbroekAJLB21} & \ding{56} & \ding{52} & \ding{52} \\

EQ5 & MSE              & $\displaystyle 
                          P(u)=P_{\text{idle}}+(P_{\text{max}}-P_{\text{idle}})(2u-u^{r})$        & \cite{DBLP:conf/ccgrid/MastenbroekAJLB21,DBLP:conf/isca/FanWB07}                                           & \ding{52} & \ding{52} & \ding{52} \\

EQ6 & Asymptotic       & $\displaystyle 
                          P(u)=P_{\text{idle}}+\frac{P_{\text{max}}-P_{\text{idle}}}{2}\bigl(1+u-e^{-u/\alpha}\bigr)$ 
                                                                                                  & \cite{DBLP:conf/ccgrid/MastenbroekAJLB21}                                          & \ding{52} & \ding{52} & \ding{52} \\

EQ7 & Asymptotic DVFS  & $\displaystyle 
                          P(u)=P_{\text{idle}}+\frac{P_{\text{max}}-P_{\text{idle}}}{2}\bigl(1+u^{3}-e^{-u^{3}/\alpha}\bigr)$
                                                                                                  & \cite{DBLP:conf/ccgrid/MastenbroekAJLB21}                                          & \ding{52} & \ding{52} & \ding{52} \\
\end{tblr}
\end{table*}
\FloatBarrier

\begin{table}[ht]
\small
\centering
\caption{Power models used in this work. E=experiment (e.g., E1=experiment 1), Asym=Asymptotic. $P_{idle}$, $P_{max}$ = power used in idle and full capacity states, $\alpha$ = utilization fraction at which the host becomes asymptotic, $r$ = calibration parameter.}
\label{table:model-archive}
\begin{tblr}{
  column{3} = {r},
  column{4} = {r},
  column{5} = {r},
  column{6} = {r},
  column{7} = {r},
  column{8} = {r},
  column{9} = {r},
  hline{1-2,6,10,14,18,20} = {-}{},
}
ID  & Name            & $P_{min}$ & $P_{max}$ & $r$    & $\alpha$ & E1                         & E2                         & E3                         \\
M1  & Sqrt            & 32   & 180  & N/A  & N/A   & \ding{52} & \ding{52} & \ding{52} \\
M2  & Sqrt            & 0    & 180  & N/A  & N/A   & \ding{56} & \ding{56} & \ding{52} \\
M3  & Linear          & 32   & 180  & N/A  & N/A   & \ding{56} & \ding{52} & \ding{52} \\
M4  & Linear          & 0    & 180  & N/A  & N/A   & \ding{56} & \ding{56} & \ding{52} \\
M5  & Square          & 32   & 180  & N/A  & N/A   & \ding{56} & \ding{52} & \ding{52} \\
M6  & Square          & 0    & 180  & N/A  & N/A   & \ding{56} & \ding{56} & \ding{52} \\
M7  & Cubic           & 32   & 180  & N/A  & N/A   & \ding{56} & \ding{52} & \ding{52} \\
M8  & Cubic           & 0    & 180  & N/A  & N/A   & \ding{56} & \ding{56} & \ding{52} \\
M9  & Mse             & 32   & 180  & 10.0 & N/A   & \ding{52} & \ding{56} & \ding{56} \\
M10 & Mse             & 32   & 180  & 0.7  & N/A   & \ding{56} & \ding{52} & \ding{52} \\
M11 & Mse             & 0    & 180  & 0.7  & N/A   & \ding{56} & \ding{56} & \ding{52} \\
M12 & Asym      & 32   & 180  & N/A  & 0.30  & \ding{52} & \ding{56} & \ding{56} \\
M13 & Asym      & 32   & 180  & N/A  & 0.85  & \ding{56} & \ding{52} & \ding{52} \\
M14 & Asym      & 0    & 180  & N/A  & 0.85  & \ding{56} & \ding{56} & \ding{52} \\
M15 & AsymD & 32   & 180  & N/A  & 0.30  & \ding{52} & \ding{56} & \ding{52} \\
M16 & AsymD & 32   & 180  & N/A  & 0.85  & \ding{56} & \ding{52} & \ding{52} \\
M17 & AsymD & 0    & 180  & N/A  & 1.90  & \ding{56} & \ding{56} & \ding{52} \\
M18 & AsymD & 32   & 180  & N/A  & 1.90  & \ding{56} & \ding{52} & \ding{52} \\
\end{tblr}
\end{table}

In \Cref{table:model-formulas}, we give an overview of the models used to predict the energy usage of the infrastructure under workload, the formulas used by the models, relevant work embedding these models, and the experiments where they have been used. In \Cref{table:model-archive}, we give an overview of the model configuration used throughout the experimentation process.

\section{Experiment 1 - performance analysis}\label{sec:appendix:b}
\label{sec:appendix:exp1}
In this section, we present complementary technical information on the performance evaluation of the M3SA system. In \Cref{sec:experiments:exp1}, we run M3SA using four individual models in the simulation and analysis process. We present in \Cref{table:nfr1-addressing} the time it took OpenDC to simulate using four individual models and the time it took M3SA to leverage, compute, and plot a Multi- and Meta-Model.

We ran the experiment on a regular-user machine (i.e., not a supercomputer): a MacBook Pro 16 with an M2 Pro chip, 16 GB RAM, MacOS Sequoia 15.0.1, and no user activities running in the background. We ran each experiment 10 times without changing the experimental setup. Although dependent on the machine, operating system, et cetera, we expect similar proportions between the simulation time and M3SA overhead on other machines, allowing M3SA to run years of real-world datacenter operation within minutes.

\begin{table}[t]
\centering
\small
\caption{M3SA Performance Evaluation (NFR1). $\sigma$ = standard deviation, over 10 samples. RWO = real-world operational time, measured in days (d=days).}
\label{table:nfr1-addressing}
\begin{tblr}{
  cells = {r},
  row{1} = {c},
  row{2} = {c},
  cell{1}{1} = {r=2}{},
  cell{1}{2} = {r=2}{},
  cell{1}{3} = {c=2}{},
  cell{1}{5} = {c=2}{},
  cell{1}{7} = {r=2}{},
  hline{1,3,11} = {-}{},
}
Samples & RWO      & Simulation Time &                           & M3SA Overhead &                           & NFR1 \\
        &          & Time            & $\sigma$ & Time          & $\sigma$ &      \\
2,016   & 7\,d    & 5.26\,s           & 0.56             & 1.35\,s         & 0.33             & 26\,\% \\
4,032   & 14\,d   & 8.82\,s           & 1.09             & 1.74\,s         & 0.29             & 19\,\% \\
10,080  & 35\,d   & 16.59\,s          & 0.71             & 2.05\,s         & 0.21            & 12\,\% \\
20,160  & 70\,d   & 30.83\,s          & 0.82             & 3.68\,s         & 0.23             & 12\,\% \\
50,400  & 175\,d  & 73.31\,s         & 1.07             & 10.91\,s        & 0.48             & 15\,\% \\
100,800 & 350\,d  & 147.11\,s          & 1.76             & 23.71\,s        & 1.97             & 16\,\% \\
201,600 & 700\,d  & 290.40\,s         & 5.09             & 50.72\,s        & 1.82             & 17\,\% \\
403,200 & 1,400\,d & 582.76\,s         & 26.09            & 111.43\,s       & 10.31            & 19\,\% 
\end{tblr}
\end{table}

\section{Evaluating M3SA on CO2-aware workload migration}
\label{sec:appendix:exp3}\label{sec:appendix:d}

In this section, we extend \Cref{sec:experiments:exp3} and present complementary technical details on CO2-aware scheduling and workload migration using M3SA capabilities. 

We analyzed the capabilities of M3SA in CO2-aware workload scheduling and migration. In the experiment, we identified that M3SA can help C-level stakeholders with CO2-aware workload scheduling, as well as increase the prediction reliability and significantly reduce CO2 predictions. We designed and engineered a greedy CO2-aware migration algorithm, which underlies the pseudocode from \ref{lst:migration-code} and migrates the workload to the lowest CO2 intensity location, assuming no migration cost.

After establishing the experimental setup, we perform the experimental process. In Appendix~\ref{sec:appendix:exp3}, we use Marconi-22, a research-specific workload containing 8,316 jobs spanned over 30 days. We leverage the number of migrations for each month and synthesize the results in Table~\ref{appendix:table:migration-count}. Then, in Figures~\ref{fig:exp3:jan2023}-\ref{fig:exp3:dec2023}, we depict the CO2 emissions per country for each month of 2023.

Evaluating the migration count for 2023, we observe that June contains the highest number of overall migrations, while January contains the least. Despite June 2023 having the most overall migrations for all the evaluated migration granularities, we observe that the other two summer months, July and August, and the mid-autumn month of October, have a similar magnitude number of migrations. The number of migrations directly depends on the energy source and, thus, weather conditions (e.g., sun, wind, river flow). This per-month evaluation of the number of migrations helped us select June 2023 as the month to run the workloads.

\lstset{language=Python, frame=single, basicstyle=\ttfamily \scriptsize}
\label{lst:migration-code}
\begin{lstlisting}
def migrate_at_granularity(metamodels, granularity){
    current = lowest_co2_at_timestamp(metamodels, 0)
    model_length = len(metamodels[0].timestamps)
    migrations = 0

    for (i = 0; i < model_length; i++){
        if (i % granularity == 0) {
            best = lowest_co2_at_timestamp(metamodels, i)

            if (current.location == best.location) {
                current = best
                migrations++
            }
        }
    }
}
\end{lstlisting}

\begin{table}
\centering
\caption{Migration Count at various intervals, for each month of the year 2023, using ENTSOE-EU-23 carbon trace.}
\begin{tabular}{|l|r|r|r|r|r|}
\hline
Month & 15\,min & 1\,h & 4\,h & 8\,h & 24\,h \\
\hline
Jan & 4 & 4 & 4 & 2 & 2 \\
Feb & 24 & 24 & 20 & 2 & 2 \\
Mar & 73 & 73 & 33 & 11 & 3 \\
Apr & 43 & 43 & 25 & 19 & 7 \\
\hline
May & 17 & 17 & 6 & 0 & 0 \\
Jun & 112 & 112 & 69 & 35 & 11 \\
Jul & 105 & 105 & 46 & 27 & 7 \\
Aug & 112 & 112 & 58 & 28 & 8 \\
\hline
Sep & 30 & 30 & 21 & 6 & 0 \\
Oct & 112 & 111 & 53 & 32 & 8 \\
Nov & 62 & 62 & 28 & 15 & 6 \\
Dec & 53 & 53 & 16 & 8 & 3 \\
\hline
\end{tabular}
\label{appendix:table:migration-count}
\end{table}

\clearpage

\begin{figure}[ht]
    \centering
    \includegraphics[width=0.85\linewidth]{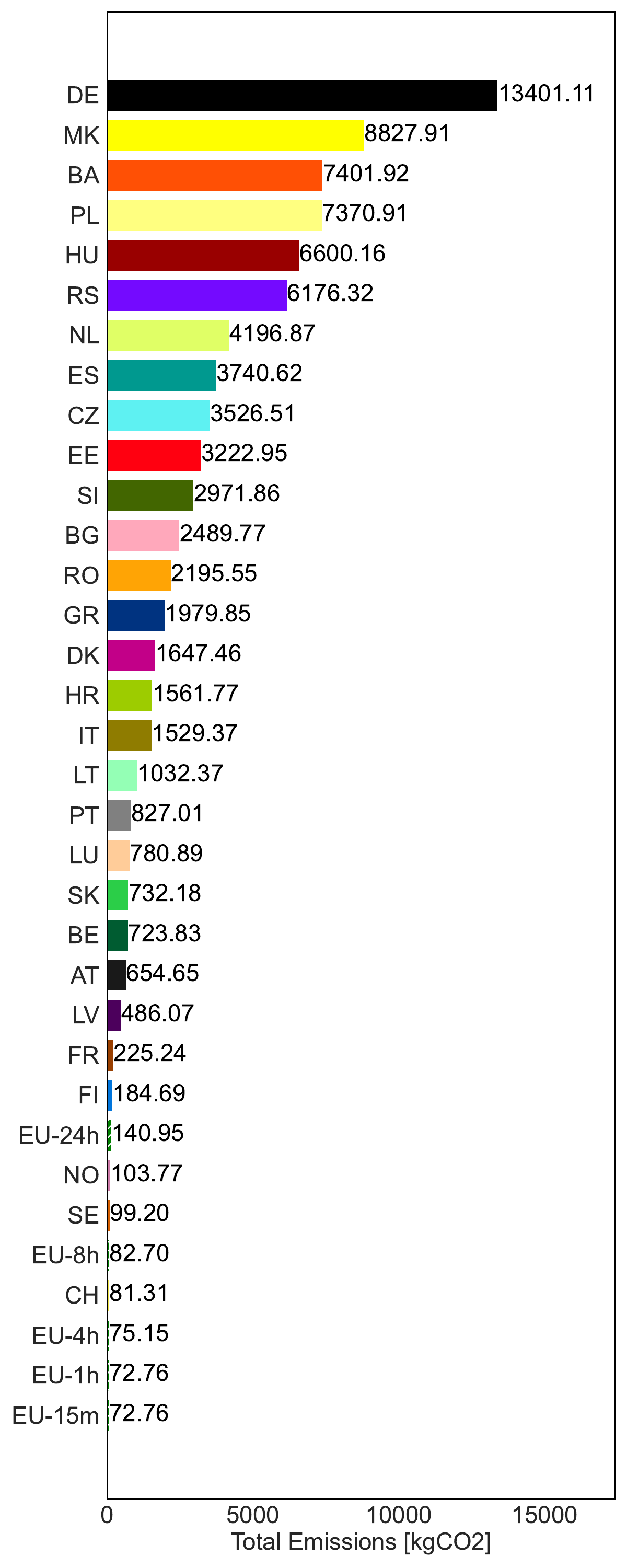} 
    \caption{Total CO2 emissions of S3 infrastructure, run in 29 different locations, and compared to 5 migration granularities. Plotted at linear scale.}
    \label{fig:exp3-all-emissions-linear-scale}
\end{figure}

\begin{figure}[ht]
    \centering
    \includegraphics[width=0.85\linewidth]{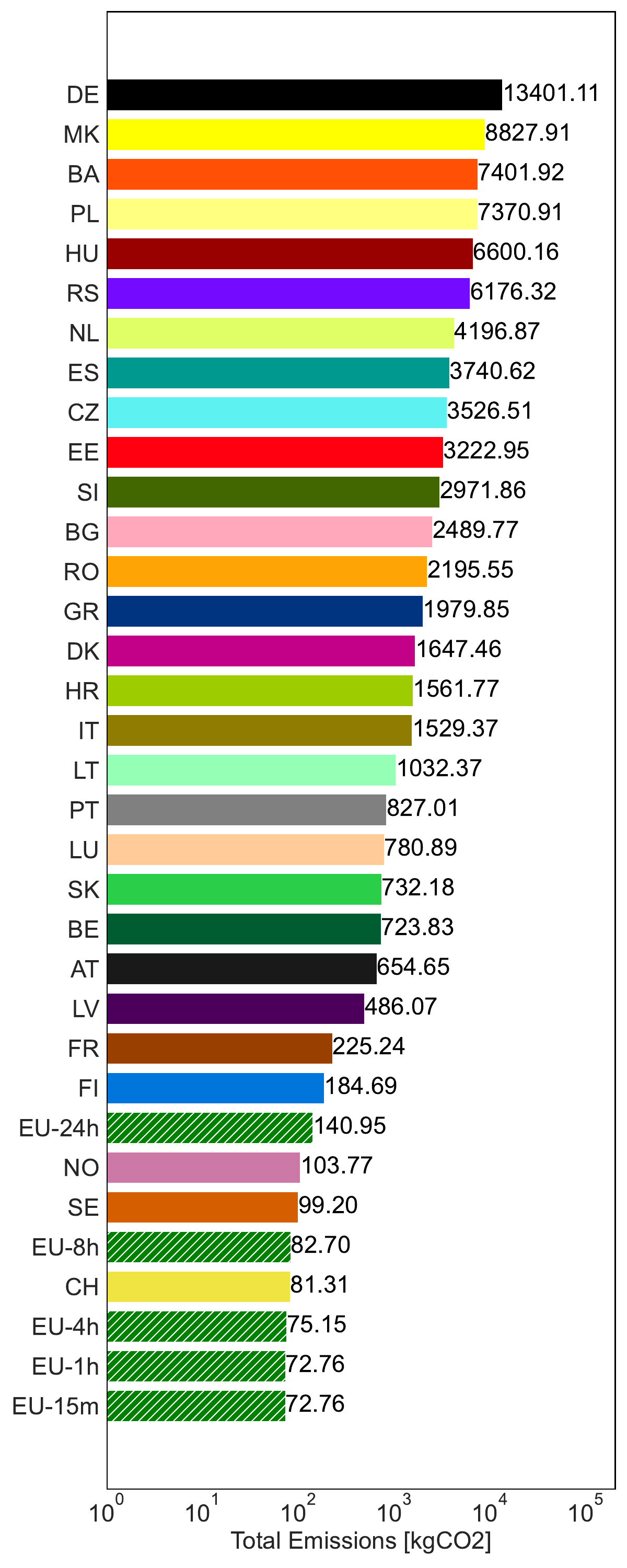} 
    \caption{Data of \Cref{fig:exp3-all-emissions-linear-scale}, plotted with logarithmic scale on the horizontal axis.}
    \label{fig:exp3-all-emissions-log-scale}
\end{figure}

\begin{figure*}
    \centering
    \includegraphics[width=0.95\linewidth]{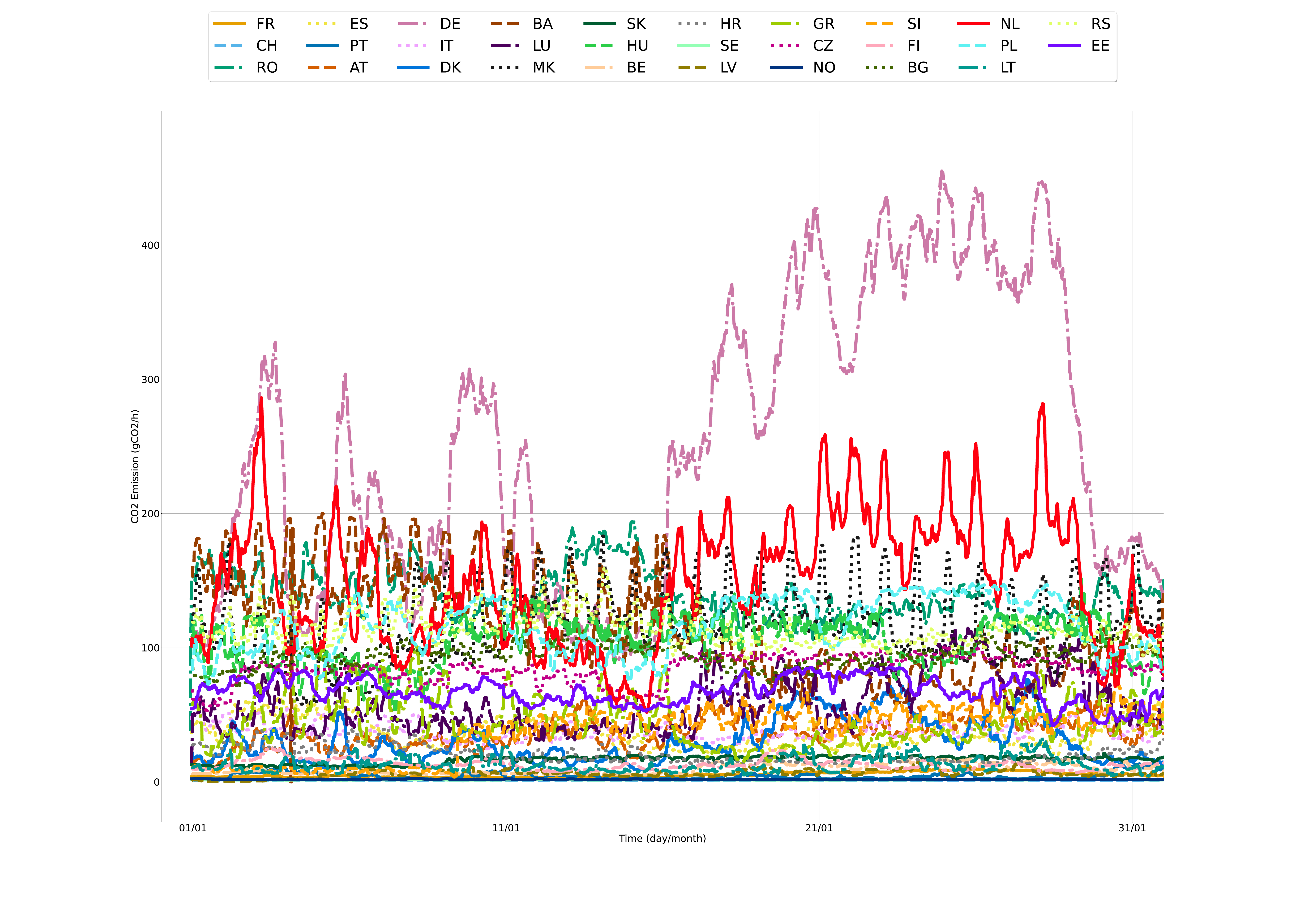}
    \caption{Legend for Figures \ref{fig:exp3:jan2023}-\ref{fig:exp3:dec2023}. The colors are kept constant for the corresponding countries.}
    \label{fig:exp3:legend}
    \vspace{-0.35cm}
\end{figure*}

\begin{figure}
    \centering
    \includegraphics[width=0.95\linewidth]{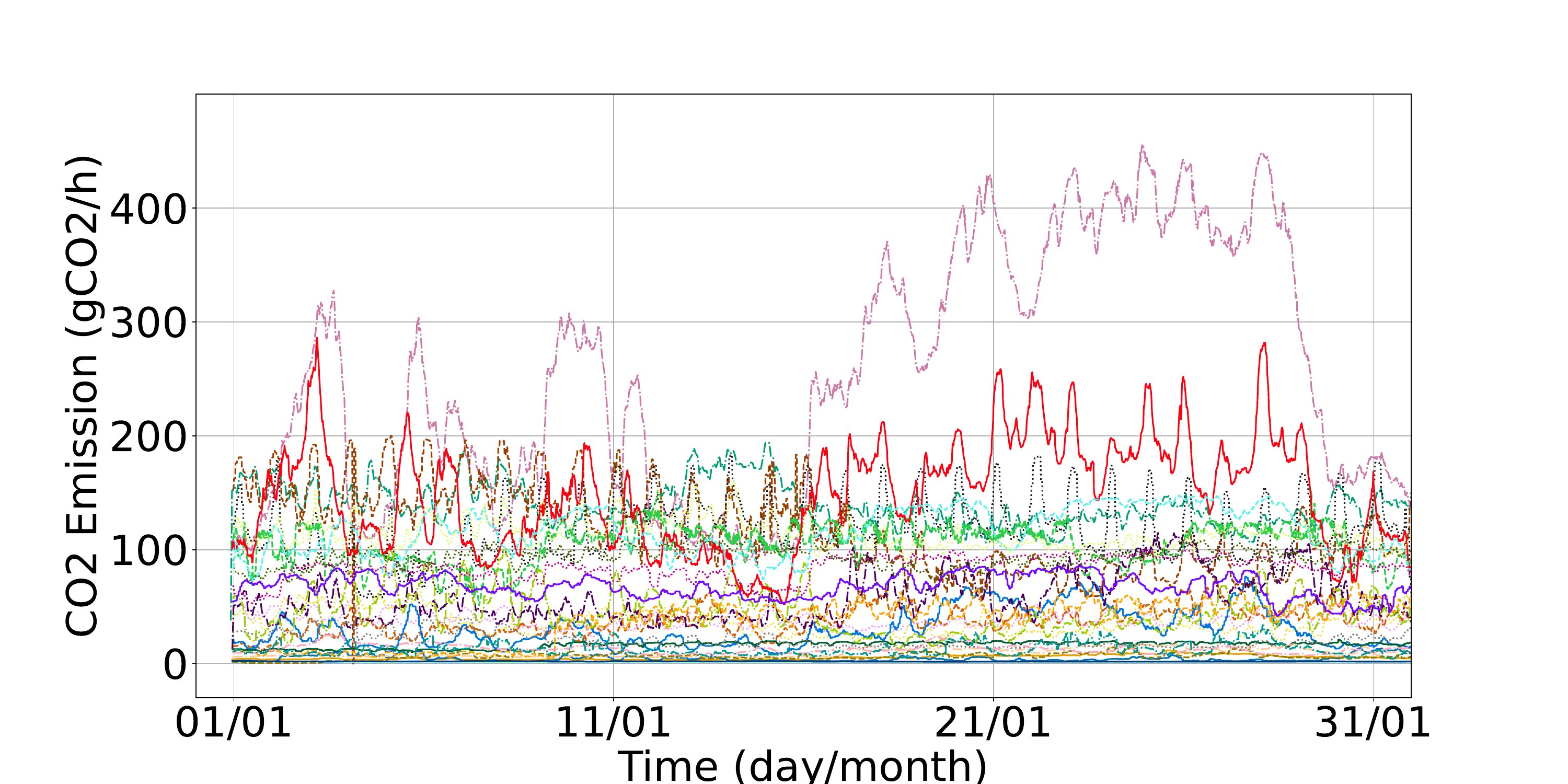}
    \caption{CO2 emissions Jan 2023.}
    \label{fig:exp3:jan2023}
    \vspace{-0.35cm}
\end{figure}

\begin{figure}
    \centering
    \includegraphics[width=0.95\linewidth]{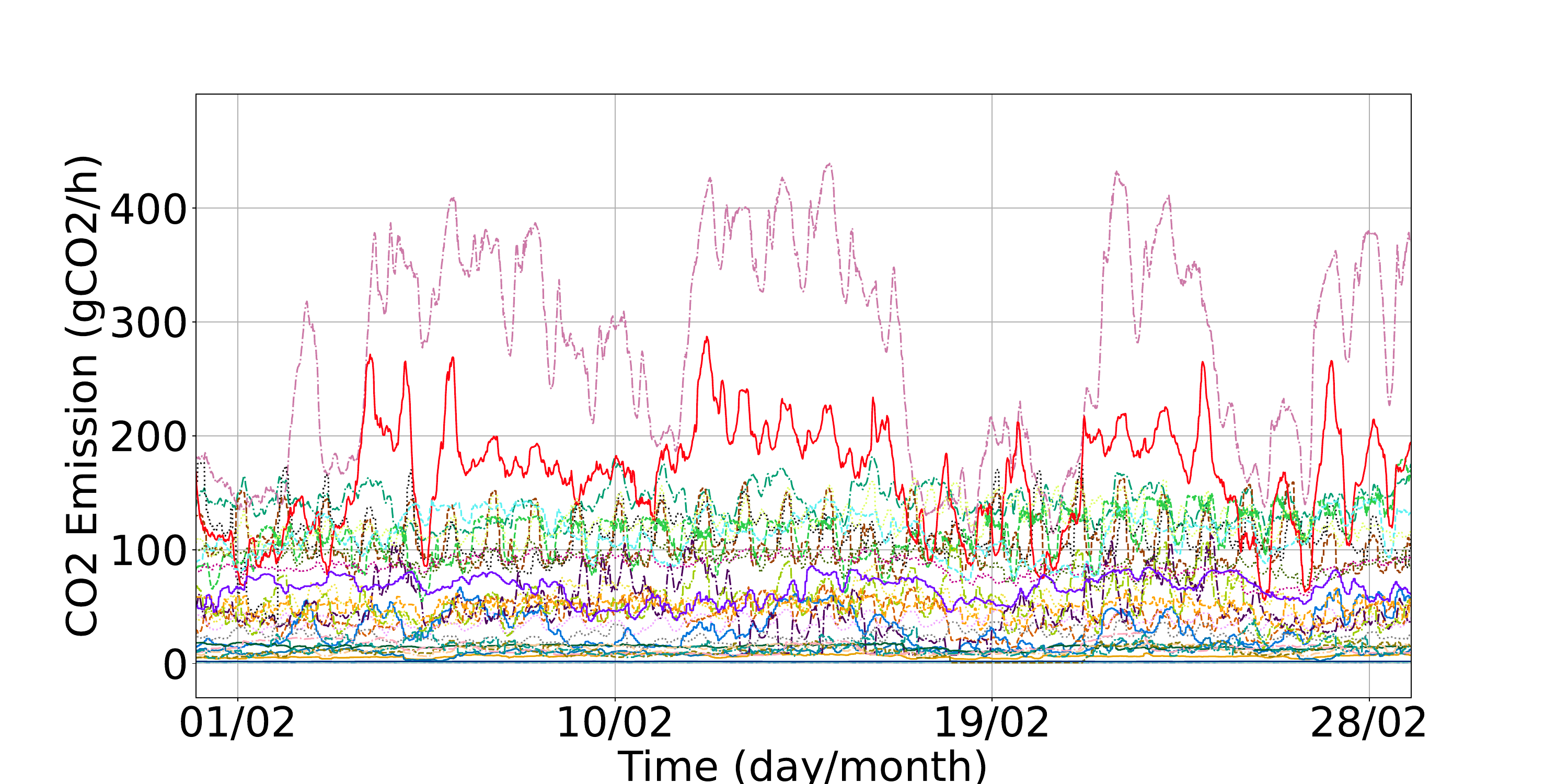}
    \caption{CO2 emissions Feb 2023.}
    \label{fig:exp3:feb2023}
    \vspace{-0.35cm} 
\end{figure}

\begin{figure}
    \centering
    \includegraphics[width=0.95\linewidth]{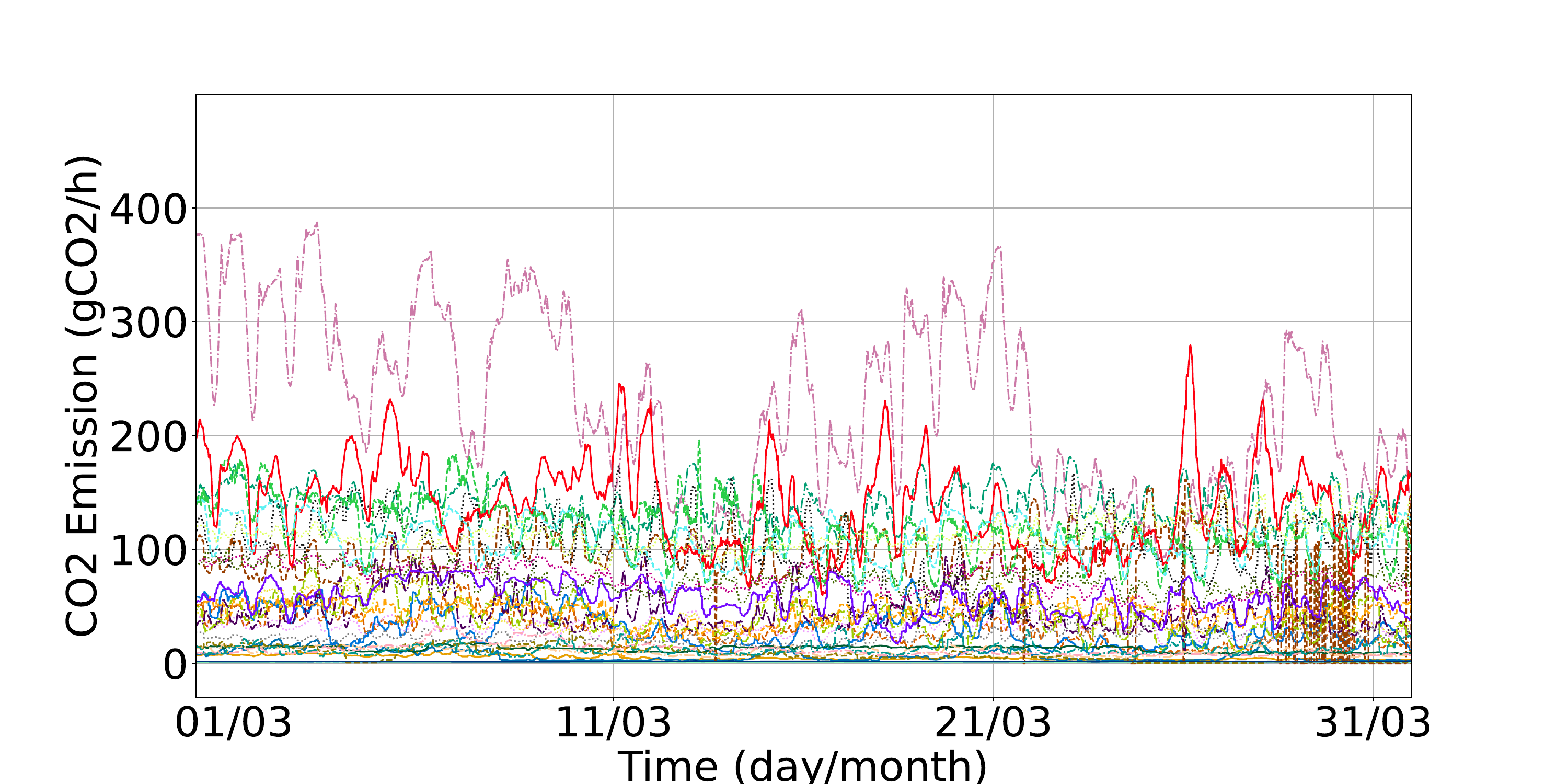}
    \caption{CO2 emissions Mar 2023.}
    \label{fig:exp3:mar2023}
    \vspace{-0.35cm} 
\end{figure}

\begin{figure}
    \centering
    \includegraphics[width=0.95\linewidth]{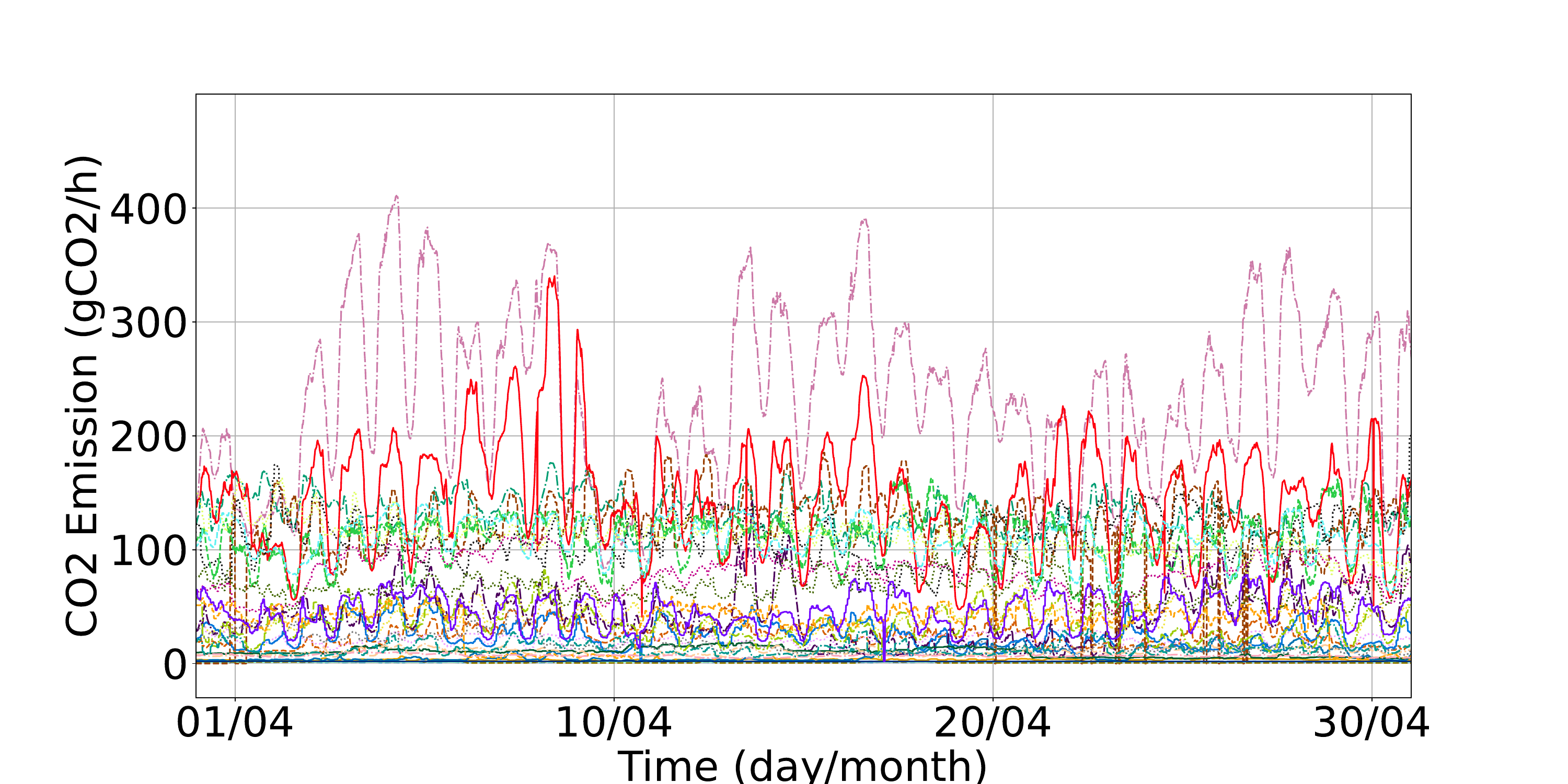}
    \caption{CO2 emissions Apr 2023.}
    \label{fig:exp3:apr2023}
    \vspace{-0.35cm} 
\end{figure}

\begin{figure}
    \centering
    \includegraphics[width=0.95\linewidth]{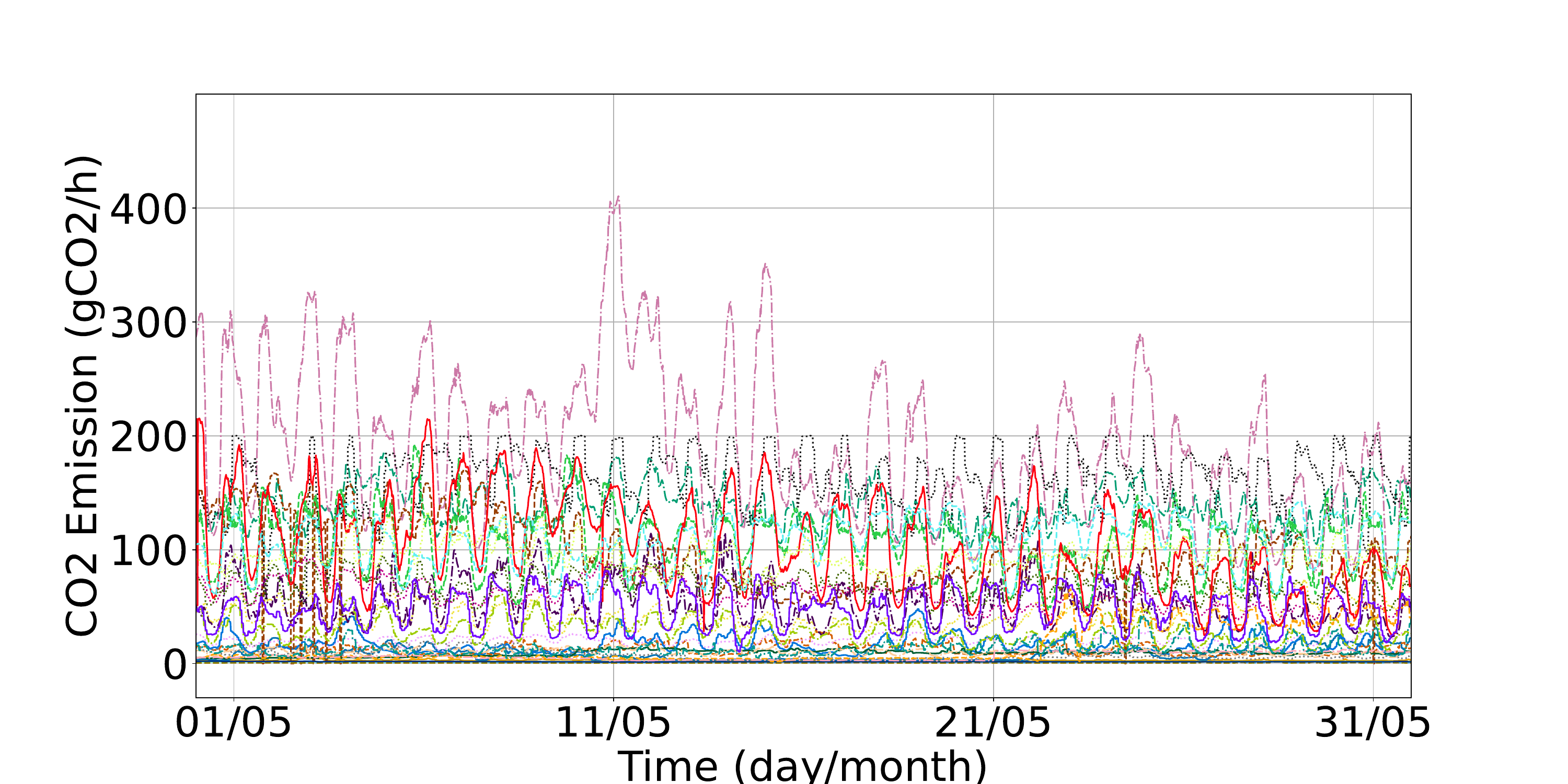}
    \caption{CO2 emissions May 2023.}
    \label{fig:exp3:may2023}
    \vspace{-0.35cm} 
\end{figure}

\begin{figure}
    \centering
    \includegraphics[width=0.95\linewidth]{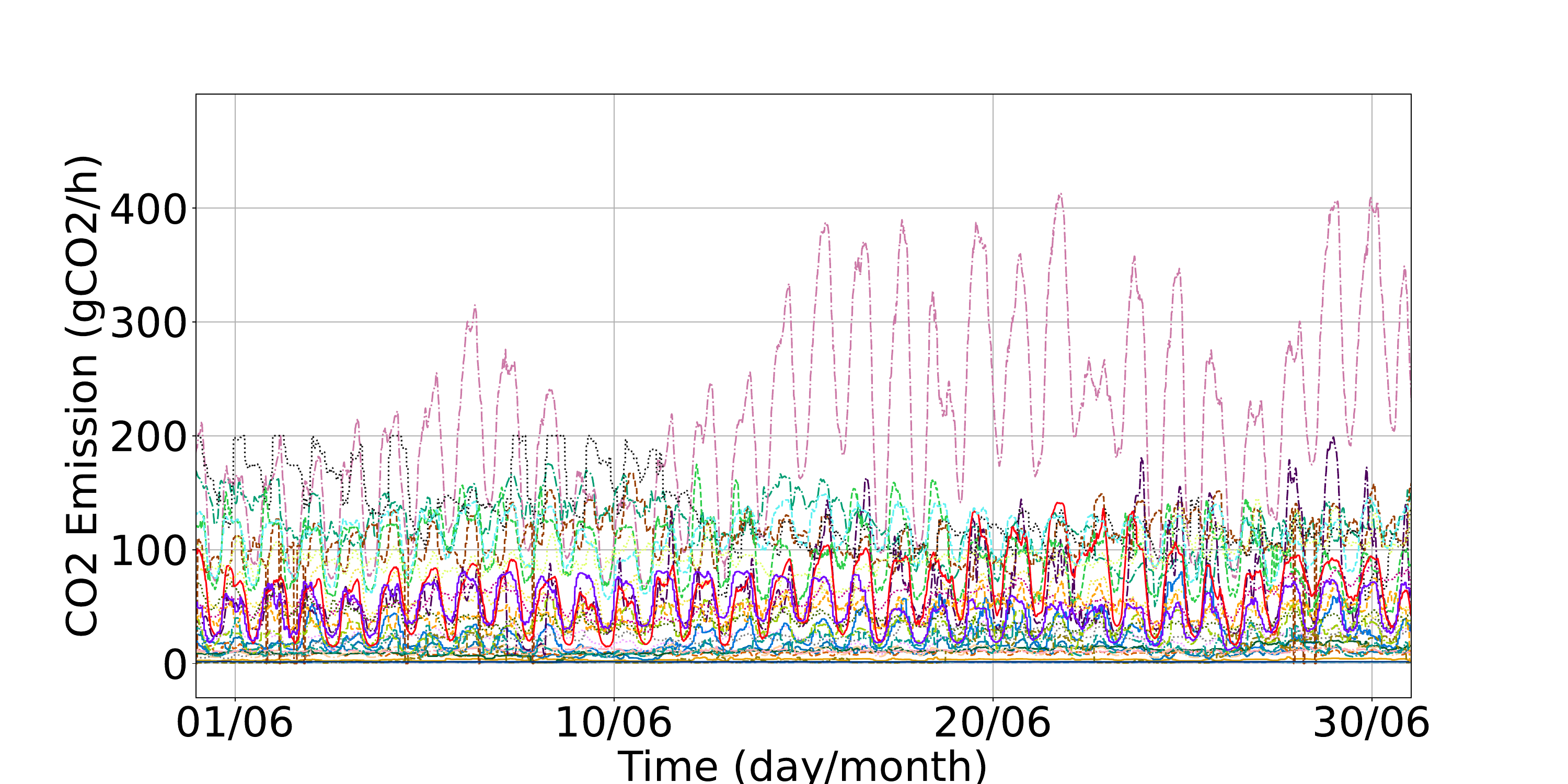}
    \caption{CO2 emissions Jun 2023.}
    \label{fig:exp3:jun2023}
    \vspace{-0.35cm} 
\end{figure}

\begin{figure}
    \centering
    \includegraphics[width=0.95\linewidth]{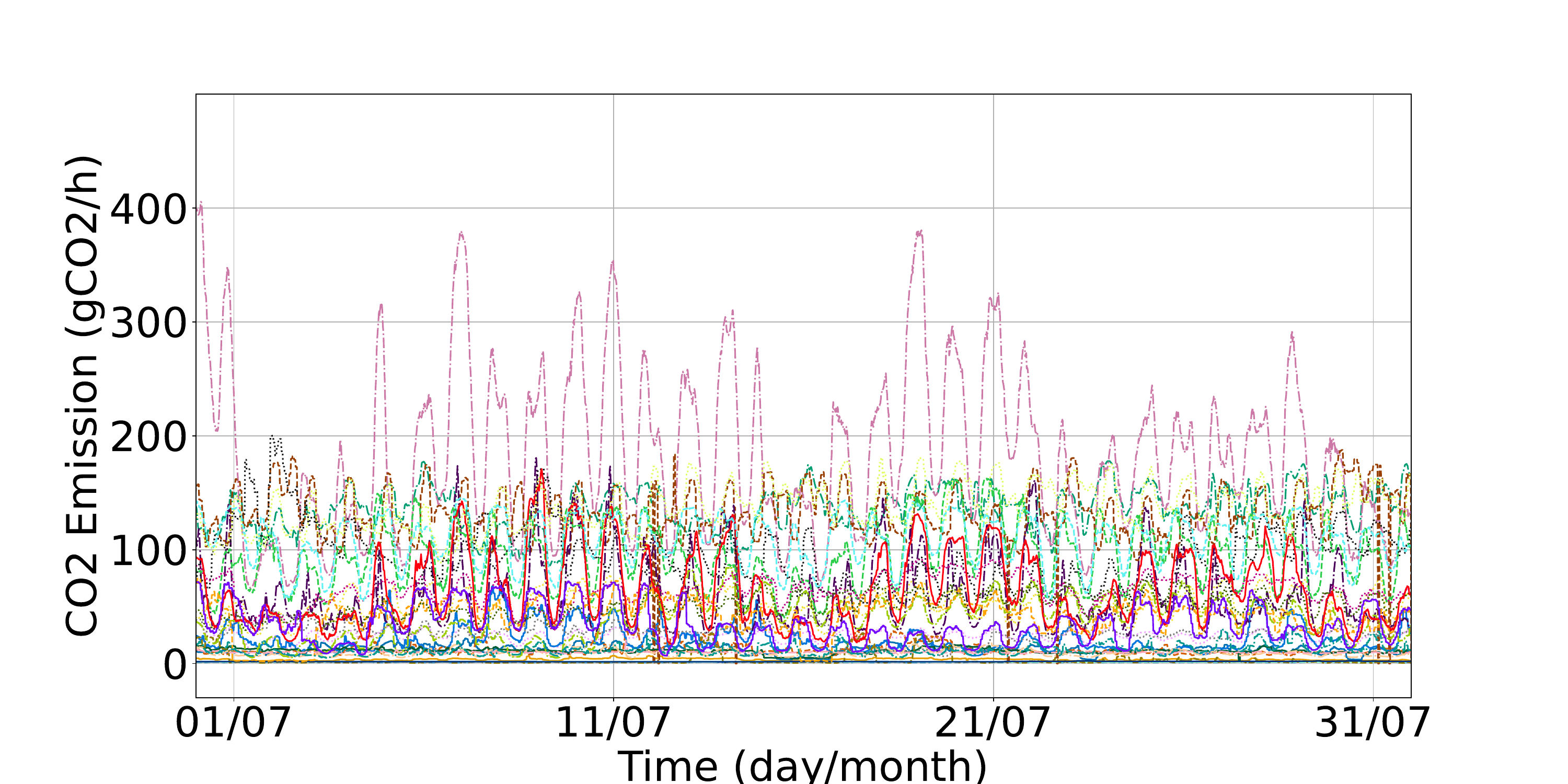}
    \caption{CO2 emissions Jul 2023.}
    \label{fig:exp3:jul2023}
    \vspace{-0.35cm} 
\end{figure}

\begin{figure}
    \centering
    \includegraphics[width=0.95\linewidth]{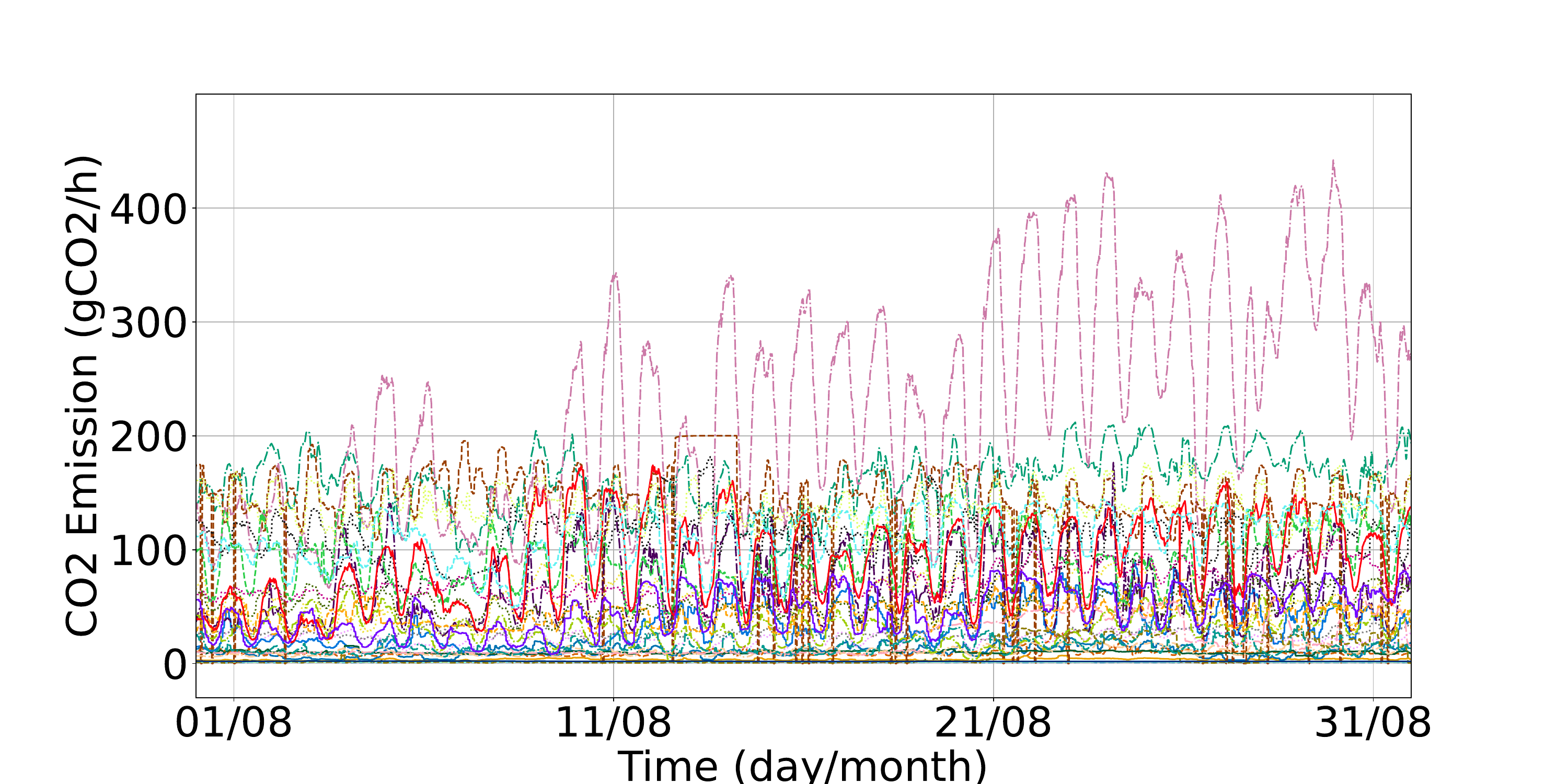}
    \caption{CO2 emissions Aug 2023.}
    \label{fig:exp3:aug2023}
    \vspace{-0.35cm} 
\end{figure}

\begin{figure}
    \centering
    \includegraphics[width=0.95\linewidth]{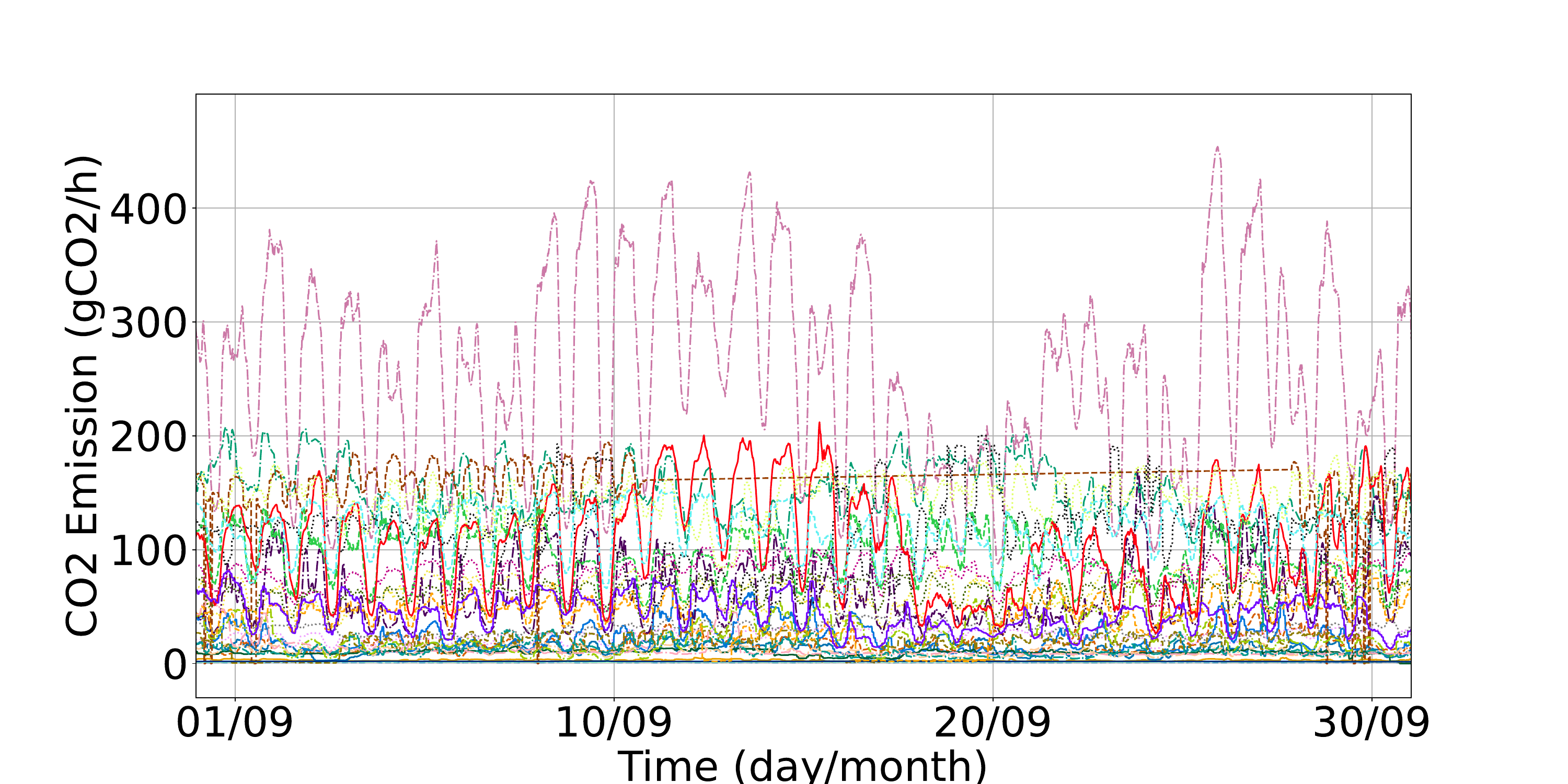}
    \caption{CO2 emissions Sep 2023.}
    \label{fig:exp3:sep2023}
    \vspace{-0.35cm} 
\end{figure}

\begin{figure}
    \centering
    \includegraphics[width=0.95\linewidth]{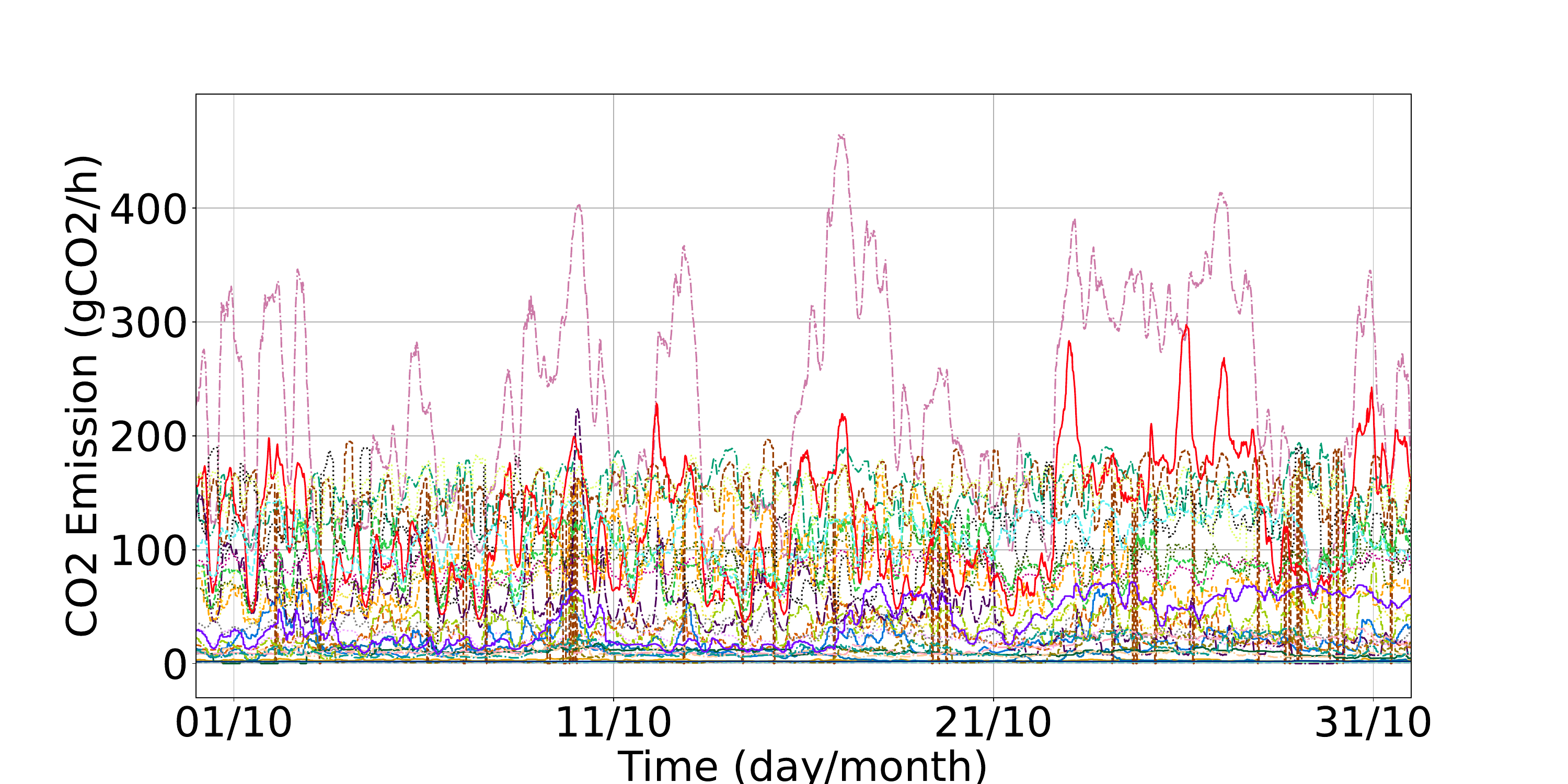}
    \caption{CO2 emissions Oct 2023.}
    \label{fig:exp3:oct2023}
    \vspace{-0.35cm} 
\end{figure}

\begin{figure}
    \centering
    \includegraphics[width=0.95\linewidth]{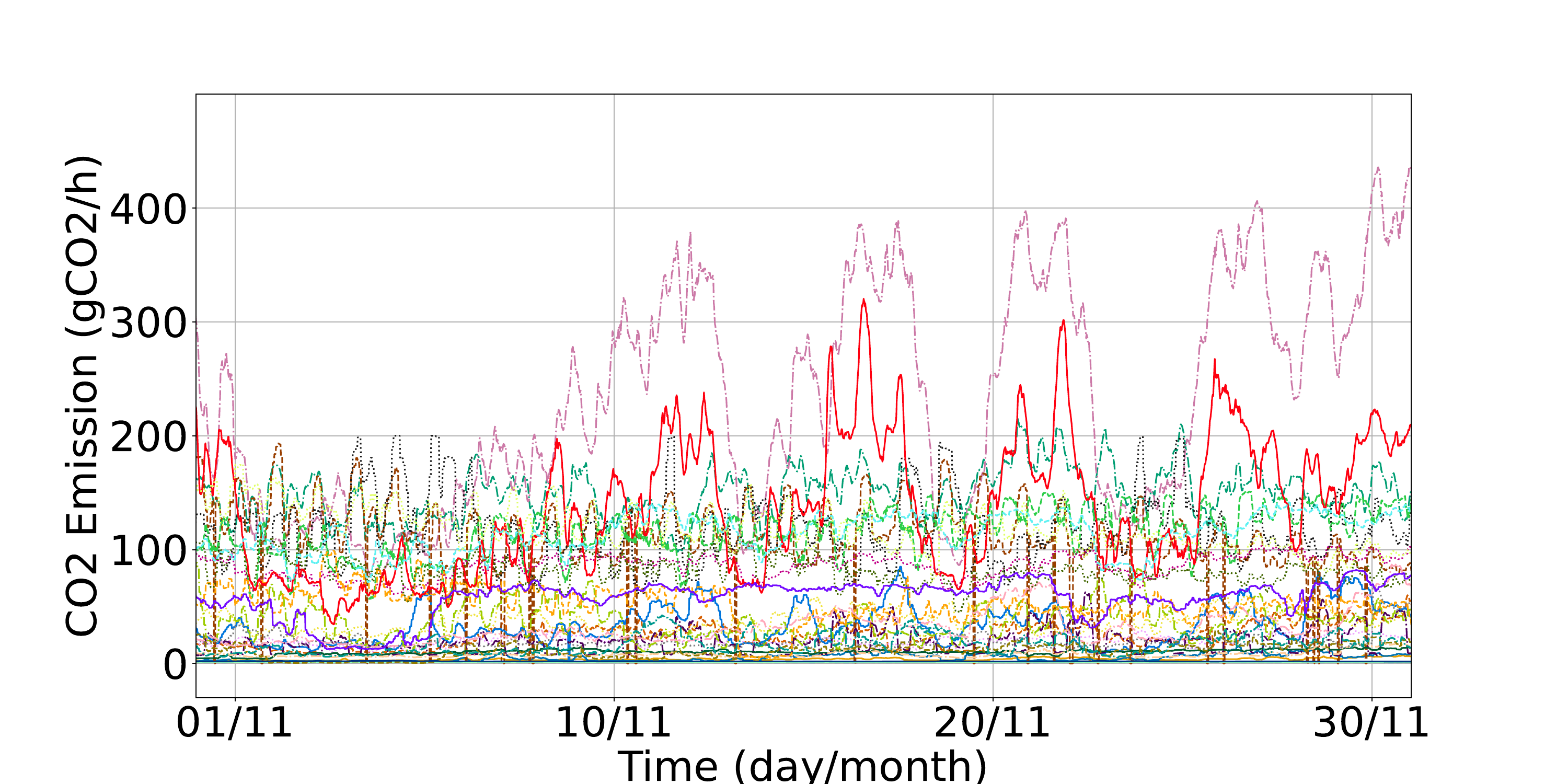}
    \caption{CO2 emissions Nov 2023.}
    \label{fig:exp3:nov2023}
    \vspace{-0.35cm} 
\end{figure}

\begin{figure}
    \centering
    \includegraphics[width=0.95\linewidth]{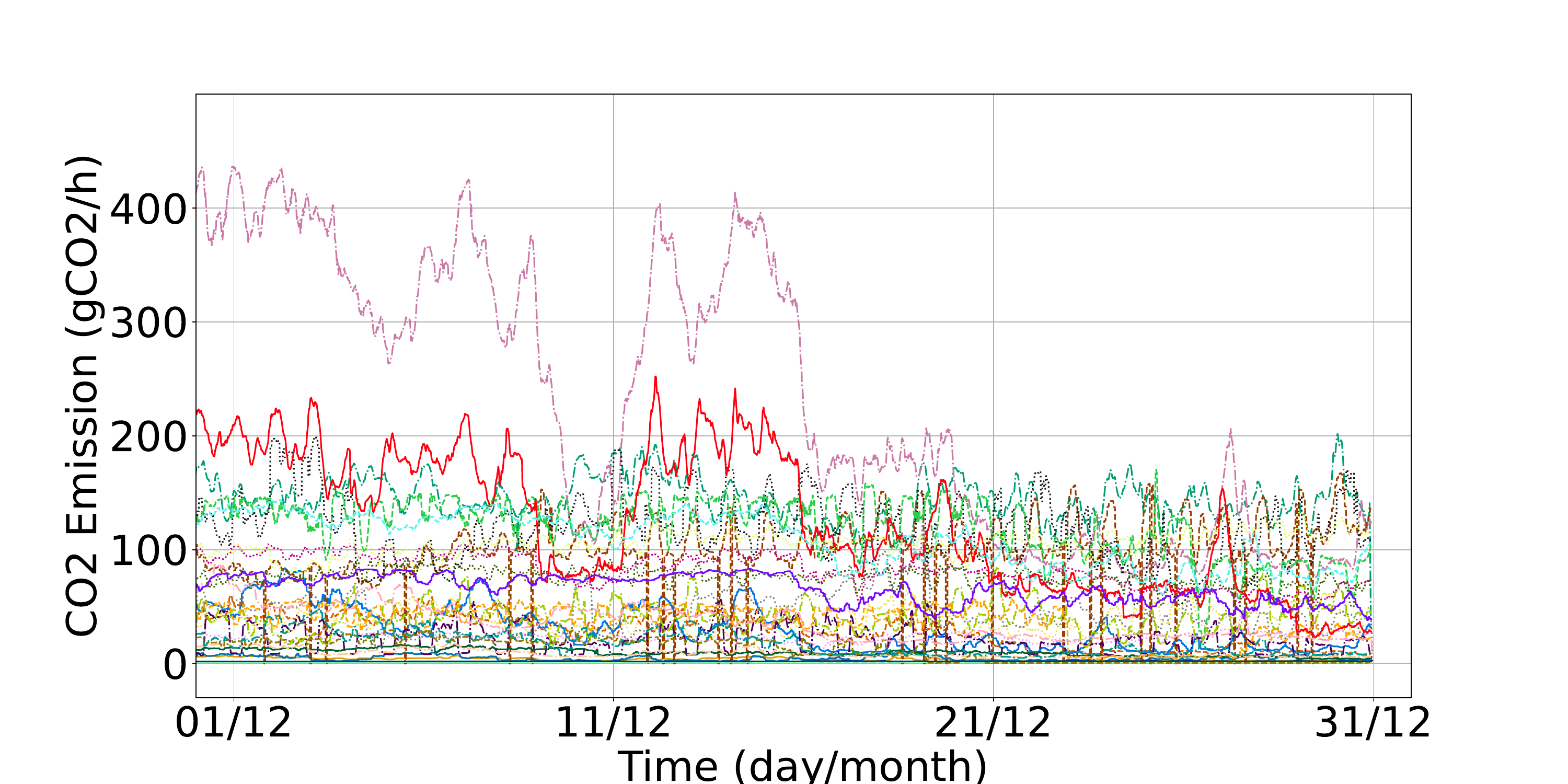}
    \caption{CO2 emissions Dec 2023.}
    \label{fig:exp3:dec2023}
    \vspace{-0.35cm} 
\end{figure}

\clearpage

\section*{D Artifact Appendix}  

\subsection*{D.1 Abstract}

The source code of M3SA, both coupled with and decoupled from OpenDC, is available at \href{https://github.com/atlarge-research/opendc-m3sa}{github.com/atlarge-research/opendc-m3sa}.
M3SA decoupled can be found in \texttt{m3sa} directory and M3SA coupled with OpenDC in \texttt{m3sa-opendc}.
We facilitate experiment and visualization reproducibility through a reproducibility capsule, available at \href{https://github.com/atlarge-research/opendc-m3sa-reproducibility-capsule}{github.com/atlarge-research/opendc-m3sa-reproducibility-capsule}. The reproducibility capsule is Docker-based and supports two main commands, one for reproducing the isolated experiment presented in the article, and a command for reproducing all the experiments and visualizations presented in the homonymous technical report.



\subsection*{D.2 Artifact check-list (meta-information)}

We present the artifact checklist for reproducing the experiment in this article. 


{\small
\begin{enumerate}
  \item {\textbf{Program:} OpenDC-M3SA}
  \item {\textbf{Binary:} Default OpenDC JAR, also provided in the reproducibility capsule}
  \item {\textbf{Data set:} workload traces available in \texttt{experiments} folder. All workloads used are available, and detailed, at \url{https://github.com/atlarge-research/opendc-traces}}
  \item {\textbf{Run-time environment:} Docker container (tested on \texttt{docker-27.4} and above, \texttt{python:3.12-slim} base image}, bundling \texttt{JDK-21})
  \item {\textbf{Metrics:} power draw in the simulated environment (in the technical report, also CO2 emissions)}
  \item {\textbf{Output:} the figures, in pdf format. Running the capsule generates (reproduces) \textit{(i) the figures from the article}, and (ii) other figures, from other experiments included in the technical report. For (i), the output figures \texttt{figure-9A.pdf}, \texttt{figure-9B.pdf}, and \texttt{figure-9C.pdf} correspond to the article's Figure~4~A, B, and C, respectively.
  }
  \item {\textbf{How much disk space is required (approximately)?:} 2.5\,GiB, including the article and technical report results.}
  \item {\textbf{How much time is needed to prepare workflow (approximately)?:} Under 5 minutes}
  \item {\textbf{How much time is needed to complete experiments (approximately)?:} $\approx$ 2~min for the experiment in the article. $\approx$ 15~min for all the capsule, including the experiments in the technical report. }
  \item {\textbf{Publicly available?:} Yes}
  \item {\textbf{Code licenses (if publicly available)?:} AGPL-3.0 }
  \item {\textbf{Data licenses (if publicly available)?:} AGPL-3.0 }
  \item {\textbf{Archived (provide DOI)?:} \href{https://zenodo.org/records/19072789}{10.5281/zenodo.19072789}}
\end{enumerate}
}

\subsection*{D.3 Description}

\subsubsection*{D.3.1 How to access}

We provide the reproducibility capsule via GitHub and Zenodo. We recommend viewing the capsule via GitHub and following the steps from the README.md.

\textit{Please note:} For the simplest and quickest reproducibility, you do not need to download the GitHub / Zenodo repository, but only to run the commands described in Appendix A.4.

\begin{enumerate}
    \item Reproducibility capsule: \url{https://github.com/atlarge-research/opendc-m3sa-reproducibility-capsule}. The reproducibility caspule is also available via on Zenodo via \href{https://zenodo.org/records/19072789}{10.5281/zenodo.19072789}.

    \item M3SA-OpenDC: \url{https://github.com/atlarge-research/opendc-m3sa} (coupled, in the \texttt{m3sa-opendc} directory).

    \item M3SA: \url{https://github.com/atlarge-research/opendc-m3sa} (decoupled, in the \texttt{m3sa} directory).
    
\end{enumerate}

\subsubsection*{D.3.2 Hardware dependencies (no specific hardware needed)}

We tested the reproducibility capsule on X86-64 and ARM architecture, on MacOS and generic Linux distributions.

\subsubsection*{D.3.3 Software dependencies}

Docker v27.4 or higher and a Unix shell for running the commands presented in Appendix A.4.



\subsection*{D.4 Installation}

The reproducibility capsule can run via Docker or locally. We recommend the Docker-based capsule, using the pre-built Docker image. We detail the steps below.

\begin{lstlisting}[language=bash]
# Step 1: Use the pre-built Docker image
# pull the pre-built Docker image
docker pull --platform linux/amd64 \
    radu33/m3sa:m3sa-experiment
    
# tag the image for the experiment
docker tag radu33/m3sa:m3sa-experiment \
    m3sa-experiment

# Step 2: Run only the experiment in the article
docker run --platform linux/amd64 --rm \
    -v $(pwd):/app/reproduced \
    m3sa-experiment experiment1

# (Optional) Step 3: Run all the experiments 
# (the experiment from the article and from the extended technical report)
docker run --platform linux/amd64 --rm \
    -v $(pwd):/app/reproduced \
    m3sa-experiment

\end{lstlisting}


\subsection*{D.5 Evaluation and expected results}

The results are the plots included in the paper, in pdf format, and are included in the current working directory.

\begin{figure}[!h]
    \centering
    \includegraphics[width=0.65\linewidth]{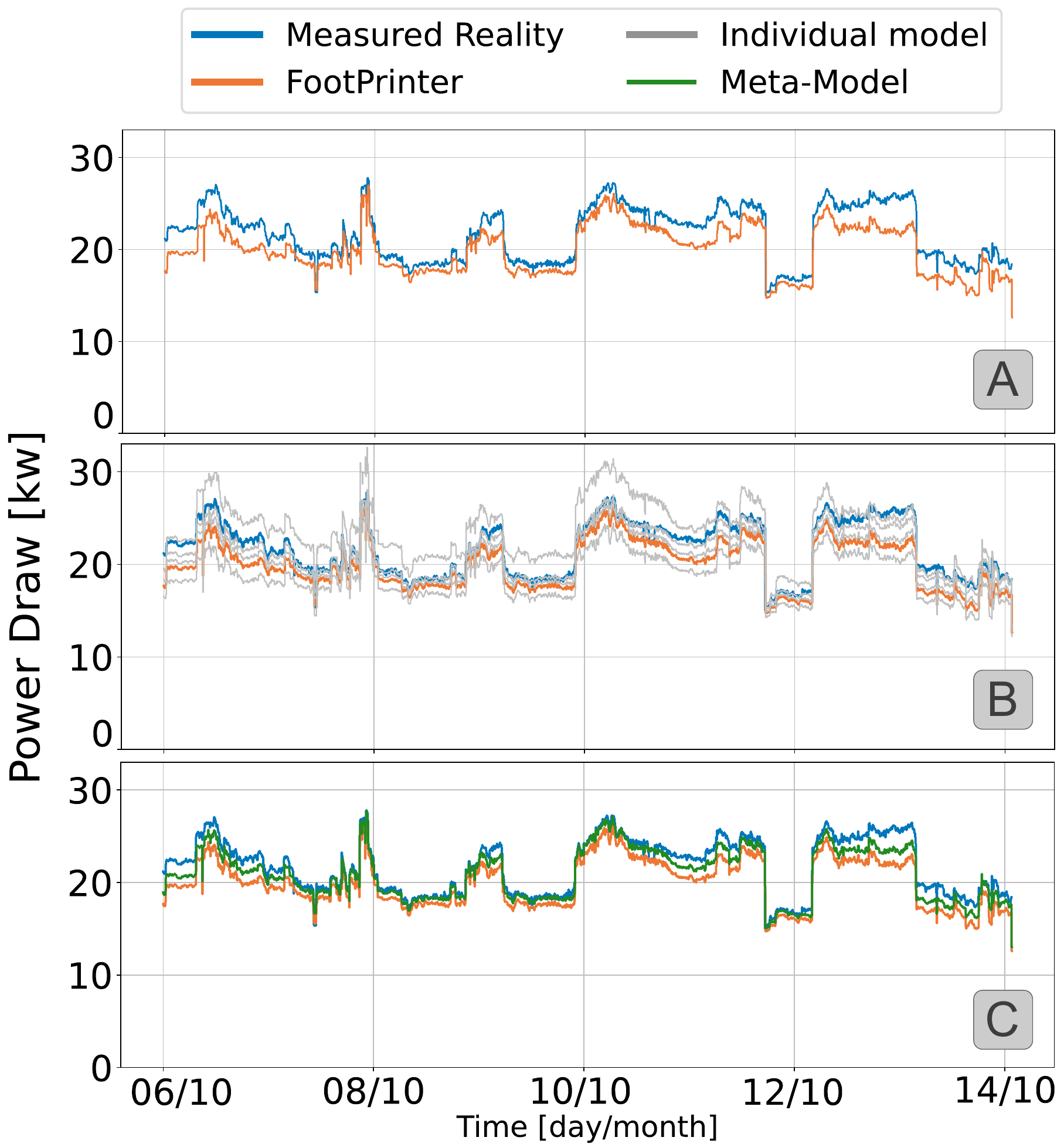}
    \caption{Figure 4A, 4B, 4C from the article, named ``figure-9A'' (-9B, -9C) in the capsule.}
    \label{fig:placeholder}
\end{figure}

The output consists of the figures in PDF format. Running the capsule generates (reproduces) \textit{(i) the figures from the article}, and (ii) other figures, from other experiments included in the technical report. For (i), the output figures \texttt{figure-9A.pdf}, \texttt{figure-9B.pdf}, and \texttt{figure-9C.pdf} correspond to the article's Figure~4~A, B, and C, respectively.







\end{document}